\documentclass[rmp,aps,eqsecnum,amsmath,amssymb]{revtex4}

\usepackage{graphics}
\usepackage{epsfig}

\def\inbar{\,\vrule height1.5ex width.4pt depth0pt}
\def\IR{\relax{\rm I\kern-.18em R}}
\def\IC{\relax\hbox{$\inbar\kern-.3em{\rm C}$}}

\begin{document}
\title{Andreev - Saint James reflections: a probe of cuprate superconductors}

\author{Guy Deutscher}
\email{guyde@tau.ac.il}

\affiliation{School of Physics and Astronomy, Sackler Faculty of
Exact Sciences Tel Aviv University, Ramat Aviv, Tel Aviv 69978,
Israel}

\begin{abstract}
Electrical transport through a normal metal / superconductor
contact at biases smaller than the energy gap can occur via the
reflection of an electron as a hole of opposite wave vector. The
same mechanism of electron-hole reflection gives rise to low
energy states at the surface of unconventional superconductors
having nodes in their order parameter. The occurrence of
electron-hole reflections at normal metal / superconductor
interfaces was predicted independently by Saint James and de
Gennes and by Andreev, and their spectroscopic features discussed
in detail by Saint James in the early sixties. They are generally
called Andreev reflections but, for that reason, we call them
Andreev - Saint James (ASJ) reflections. We present a historical
review of ASJ reflections and spectroscopy in conventional
superconductors, and review their application to the High $T_c$
cuprates. The occurrence of ASJ reflections in all studied
cuprates is well documented for a broad range of doping levels,
implying that there is no large asymmetry between electrons and
holes near the Fermi level in the superconducting state. In the
underdoped regime, where the pseudo-gap phenomenon has been
observed by other methods such as NMR, ARPES and Giaever
tunneling, gap values obtained from ASJ spectroscopy are smaller
than pseudo-gap values, indicating a lack of coherence in the
pseudo-gap energy range. Low energy surface bound states have been
observed in all studied hole doped cuprates, in agreement with a
dominant d-wave symmetry order parameter. Results are mixed for
electron doped cuprates. In overdoped $YBa_2Cu_3O_{7-\delta}
(\delta<0.08)$ and $La_{2-x}Sr_xCuO_4$, ASJ spectroscopy is
consistent with the presence of an additional imaginary component
of the order parameter. Results of ASJ spectroscopy under applied
magnetic fields are also reviewed. A short section at the end is
devoted to some recent results on spin effects.
\end{abstract}

%\date{May 2001}
\maketitle

\newpage
\tableofcontents
\newpage
\section{INTRODUCTION}
\label{sec:intro} When an electron moving in a normal metal N with
momentum k hits an interface with a superconductor S, it is
reflected as a hole of equal momentum if its kinetic energy
measured from the Fermi level is smaller than the energy gap
$\Delta$ of S. Because of its negative effective mass, the
reflected hole has a velocity opposite to that of the incoming
electron, and carries charge current in the same direction. This
process, known to day as an Andreev reflection \cite{Andreev:1964}, was
in fact first described by de Gennes and Saint James \cite{De_Gennes:1963} (and
in a follow up paper by Saint James \cite{Saint_James:1964}) . The two independent
original papers show solutions of the Bogoliubov -de Gennes
equations \cite{De_Gennes:1966}, based on the Bogolubov transformation
\cite{Bogoliubov:1958}, for the pair potential near an N/S interface.
But they focus on different phenomena.

Andreev was interested in heat transport in the intermediate state
of a Type I superconductor, and showed that domain walls provide a
resistance to this flow because of the electron - hole reflection
mechanism. This is why the thermal resistance in the intermediate
state is higher than in the Meissner state, as observed by
Mendelssohn and Olsen \cite{Mendelssohn:1950}. Andreev was able to fit
quantitatively the detailed heat transport measurements of
Zavaritskii \cite{Zavaritskii:1960}. De Gennes and Saint James, on the other hand,
were interested in the Density of States (DOS) in a system
consisting of a normal slab of thickness $d_N$ in close electrical
contact with a semi-infinite superconductor. They showed that the
DOS has a series of peaks below $\Delta$ because of the existence
of finite energy bound states in N. For large enough values of
$d_N$, these peaks are located at energies that are multiple of
($\hbar v_F/4 d_N$), where $v_F$ is the Fermi velocity in N. Saint
James \cite{Saint_James:1964} remarks that the inter-level distance is half of that
for an electron in a potential well, and explains that this is
because a complete cycle comprises two electron-hole reflections
at the N/S interface and two specular reflections at the outer
surface of N. In this cycle, the quasi-particle has both an
electron like and a hole like character.

The Andreev 1963 paper was and still is widely quoted, may be
because it explained a specific and observed physical phenomenon.
The slightly earlier paper of de Gennes and Saint James \cite{De_Gennes:1963} is
less quoted, and the slightly posterior paper of Saint James
\cite{Saint_James:1964} is only known to some experts in the field. It is an
excellent paper, whose reading I highly recommend. Its emphasis on
the spectroscopic aspects of the electron-hole reflections is very
close to our current interest, which will focus here on the use of
these reflections for the spectroscopic study of High Temperature
Superconductors (HTSC). After consulting with some colleagues, I
decide for this reason to use in this review the terms of Andreev
- Saint James (ASJ) reflections and spectroscopy.

Another curious aspect of the history of ASJ reflections is that
it took almost 20 years before it was shown theoretically that
they \emph{enhance} the electrical conductance of N/S contacts at
biases below the gap. This is in contrast with the
\emph{reduction} of the thermal conductance. This enhancement of
the electrical conductance appears obvious to us to day. An
electron coming in from the N side at energies smaller than
$\Delta$ cannot propagate inside S, only Cooper pairs. The
reflected hole ensures current conservation. A charge of $2e$ then
flows across the interface, which corresponds to an increase of
the conductance of the contact by a factor of 2 compared to that
in the normal state, or at biases much larger than the gap. The
detailed way in which the process occurs involves the creation of
electron and hole excitations in S near the interface, which
recombine into pairs over the coherence length of the
superconductor. Pankove \cite{Pankove:1966} did report an enhanced electrical
conductance of N/S contacts below the gap. His observations on
pressure contacts between Al and Nb are clear: "When a contact is
made between a normal metal and a superconductor, the V-I
characteristic of the contact shows an initial region of high
conductance with an abrupt change to a region of lower
conductance". However, Pankove did not relate his observation to
the works of Andreev and de Gennes and Saint James. Likewise, his
observations were not noted by theorists. Griffin and Demers
\cite{Griffin:1971} were the first to calculate the  quasi-particle
transmission probability for \emph{excitations} going from a
normal to a superconducting region for N/S contacts of various
transparencies, but did not take into consideration the
electron-hole reflection mechanism and the corresponding flow of
\emph{pairs}. It is not until 1980 that Zaitsev \cite{Zaitsev:1980}
calculates an enhanced conductance below the gap, and only in 1982
Blonder, Tinkham and Kalpwijk (BTK) \cite{Blonder:1982} give a complete theoretical
discussion, including the effect of an imperfect (not fully
transparent) interface, and successfully compare their predictions
to measurements performed on Point Contacts.

Sharvin \cite{Sharvin:1965} had been the first to note that the electrical
resistance of an ideal intermetallic contact of a size smaller
than the electronic mean free path is determined by the number of
quantum channels through the contact. These contacts, called
Sharvin contacts or Point Contacts are the ideal tool for the
study of ASJ reflections, for reasons that will be reviewed later
in great detail. However, Sharvin apparently did not use them for
that purpose, another strange twist in the history of ASJ
reflections. Pankove contacts were actually Sharvin contacts, he
calculates that their size is of about $50\AA$, a typical Point
contact contact size.

The transition between the region of higher conductance and that
of lower conductance noted by Pankove occurs when the bias across
the contact is equal to the energy gap. ASJ reflections are
therefore a useful tool for the determination of the gap. BTK made
this tool a quantitative one by taking into account a non ideal
nature of the contact, including the effect of a thin insulating
barrier and that of a mismatch of the Fermi velocities between the
two metals. Yet, their work had only a limited impact on the study
of Low Temperature Superconductors (LTSC), may be because it came
much later than the full development of tunneling theory
\cite{McMillan:1969} and also, I suppose, because the controlled
fabrication of thin tunneling dielectric barriers had been
achieved so successfully following the work of Giaever
\cite{Giaever:1960}.

By contrast, ASJ reflections have become a major tool for the
study of unconventional superconductors,such as heavy fermions
\cite{Hasselbach:1993,Goll:1993}, organic superconductors (see for
instance \cite{Ernst:1994}) and HTSC, which are the focus of this
review. This development is the object of this review.  One of the
reasons why we became involved in this field was that we had
strong doubts that dielectric junctions of a quality comparable to
that achieved by Giaever on LTSC could ever be achieved on HTSC.
Dielectric junctions made on LTSC are based on the oxidation of
the metal. This method is not applicable to HTSC, because they are
oxides by themselves, and further oxidation renders them even more
metallic (and eventually non superconductors). Further, the
fabrication of HTSC requires high temperatures of the order of 700
to 800 degrees Centigrade and the use of single crystal
substrates, which precludes growing them on top of a regular metal
previously oxidized. In our laboratory, we therefore decided to
concentrate on the point contact route. In fact, observation of
ASJ reflections turned out to be relatively easy. I believe that
\cite{Hass:1992} were the first to report such observations on
single crystal quality YBCO samples, using a Gold tip as the
normal metal. For reasons that became clear only later, and that
had to do with the d-wave symmetry of the order parameter in this
superconductor, the fit to BTK theory was not perfect, but a gap
value of 18-20 meV could clearly be obtained. This value still
stands to day.

However, a determination of the energy gap is not the only, and
may be not the most important result of ASJ spectroscopy of HTSC
materials. The following points will give the reader a preliminary
idea of what the main results are:

a)  A successful quantitative fit of point-contact data to the BTK
theory, as has now been achieved in optimally doped samples, means
that the BdG equations or in other words a Fermionic description
of the excitations is appropriate for HTSC. ASJ reflections cannot
occur without electron-hole mixing.

b)  According to Blonder and Tinkham \cite{Blonder:1983}, an enhanced
conductance below the gap is only possible if the Fermi velocities
of the normal tip and of the superconductor are not too different.
That this should be the case in HTSC/ normal metal contacts is by
no means trivial. In fact, Angle Resolved Photo Emission
Spectroscopy (ARPES) data indicate for the HTSC a Fermi velocity
of the order of $2\cdot10^7$ cm/sec \cite{Margaritondo:1998}, almost
one order of magnitude smaller than that of Gold. This apparent
contradiction between experiment and theory was explained by
Deutscher and Nozieres \cite{Deutscher:1994} as resulting from a renormalization
of the Fermi velocity in the Point Contact experiments, which is
different from the full quasi-particle renormalization. This
special renormalization also explains the occurrence of strong ASJ
reflections in heavy Fermions \cite{Hasselbach:1993}.

c)  The BCS approximation of an energy gap that is very much
smaller than the Fermi energy applies extremely well to LTSC, for
which the gap value is typically less than 1 meV, and the Fermi
energy value is of several eV. It applies only marginally to the
HTSC at optimum doping, where the gap value is of a few 10 meV and
the Fermi energy value a few 100 meV. It may not apply at all in
underdoped samples, where a Bose Einstein condensation regime
could be approached. The properties of ASJ reflections in a regime
that is intermediate between BCS and Bose-Einstein condensation
\cite{Leggett:1980,Nozieres:1985} are a subject of
great current interest. In particular, the existence of ASJ
reflections in the presence of preformed pairs (or, more
generally, of a pairing amplitude) without phase coherence is
under intense consideration.

d)  ASJ reflections are phase sensitive. This major difference
with conventional Giaever tunneling spectroscopy turns out to be
of great interest for the study of superconductors having an
unconventional symmetry order parameter, as is the case for the
HTSC. As shown by Hu \cite{Hu:1994}, d-wave symmetry results in zero
energy surface bound states, or ASJ bound states, when the
orientation of the surface with respect to the crystallographic
axis is such that there are interference effects between ASJ
reflections from lobes of the order parameter of opposite signs.
This is in contrast with the finite energy bound states calculated
by de Gennes and Saint James. The case of p-wave superconductors
had been earlier investigated by Buchholz and Zwicknagl \cite{Buchholz:1981}.

This review is organized as follows. In Section II, I briefly
present the original calculation of de Gennes and Saint James for
a 1D situation giving the finite energy of bound states in a
normal slab in contact with a superconductor, and contrast it with
the zero energy states obtained in a hypothetical situation where
the normal slab is sandwiched between two superconductors whose
order parameters have phases that differ by $\pi$ . This serves as
an introduction to the effect of d-wave symmetry.

Section III is devoted to a brief summary of the BTK theory and to
Point Contact experiments in geometries where the d-wave symmetry
plays only a minor role. These experiments lead to determinations
of the gap and of a Fermi velocity that is different from the
fully renormalized value. Renormalization of the later as
appropriate for point-contact spectroscopy is included in this
Section. Effects related to the d-wave symmetry, and in particular
the occurrence of surface bound states, are discussed in Section
IV. This Section includes the effect of surface currents on these
states. Section V is devoted to ASJ reflections in the underdoped
or so-called pseudogap regime, in relation to a possible BCS to
Bose Einstein condensation crossover and other pseudogap models.
Other advanced topics, such as the occurrence and detection of a
minority imaginary component ( \emph{is} or \emph{id}) of the
order parameter and the proximity effect between HTSC and normal
metals, are discussed in Section VI.

\newpage

\section{SOLUTION OF THE BdG EQUATIONS NEAR AN N/S INTERFACE}
\label{sec:II} In order to make this review self-contained, we
shall briefly outline here the main steps of the derivation that
can be found in details in Saint James \cite{Saint_James:1964}.

We consider two metals in ideal contact, the only difference
between them being that one is a superconductor and the other a
normal metal. We concentrate on quasi-particle excitations having
an energy $\varepsilon$ measured from the Fermi level smaller than
the energy gap $\Delta$. Such excitations will necessarily decay
in S. They do so over a certain length scale, which turns out to
be on the order of the coherence length $\xi$ of the
superconductor. At larger distances from the interface, all
electrons are paired. Thus, the reflection process that we have
briefly described in the introductory section, by which an
electron coming from the N side is reflected as a hole, does not
occur abruptly at the interface, but over the length scale $\xi$ .
This property, which appears explicitly in the calculation, will
be of great importance when we discuss contacts with HTSC, in
particular in the so called pseudogap region.

As it is, the calculation ignores the effect of the proximity of N
on the value of the gap $\Delta$ in S near the interface. It is
well known that this effect cannot in general be neglected and
that a depression of $\Delta$ occurs over the length $\xi$
\cite{Deutscher:1969}. However, in the actual Point
Contact set up described by BTK (see next Section) the contact
size is smaller than $\xi$, so that this depression effect is much
reduced and can be neglected.

\subsection{The case of a normal slab in contact with a semi-infinite superconductor}
\label{sec:IIA}

In a normal metal, the excitation energies of electrons and holes
are derived respectively from the following relations:
\begin{eqnarray}
\label{eq:e21a} \varepsilon u = [-(\hbar^2/2m)\nabla -E_F]u
 \\
 \label{eq:e21b} \varepsilon v = [(\hbar^2/2m)\nabla +E_F]v
\end{eqnarray}
In a superconductor, excitations have a mixed electron-hole
character and the above equations are complemented by crossed
terms:
%\begin{subequations}
\begin{eqnarray}
\label{eq:e22a} \varepsilon u = [-(\hbar^2/2m)\nabla -E_F]u +
\Delta v
 \\
\label{eq:e22b} \varepsilon v = [(\hbar^2/2m)\nabla +E_F]v +
\Delta^* u
\end{eqnarray}
%\end{subequations}

 The solutions for u and v are oscillatory in N,
and for $\varepsilon < \Delta$  are decaying in S with complex
wave vectors. The conditions at the interface, assuming that it is
perfectly transparent, are that $u$ and $v$ and their derivatives
are continuous. At the outer surface of N, assumed to be bonded by
a dielectric, the boundary conditions are that u and v are zero.
With the normal slab having a thickness $d_N$ , and taking the
origin at the interface, one dimensional solutions in N are of the
form:
\begin{eqnarray}
\label{eq:e23a} u = \alpha sin[k'_1 (x+d_N)]
 \\
\label{eq:e23b} v = \beta sin[k'_2 (x+d_N)]
\end{eqnarray}
with :
\begin{eqnarray}
\label{eq:e24a}    k'_1=[(2m/\hbar^2)(E_F + \varepsilon)]^{1/2}
 \\
\label{eq:e24b}    k'_2=[(2m/\hbar^2)(E_F - \varepsilon)]^{1/2}
\end{eqnarray}
Solutions in S are of the form:
\begin{eqnarray}
\label{eq:e25a} u = \alpha_1 exp(ik_{1}x) + \alpha_2 exp(-ik_{2}x)
 \\  \label{eq:e25b} v = \beta_1 exp(ik_{1}x) + \beta_2 exp(-ik_{2}x)
\end{eqnarray}
The wave vectors are given by:
\begin{eqnarray}
\label{eq:e26a}
k_1=(2m/\hbar^2)^{1/2}[E_F+(\varepsilon^2-\Delta^2)^{1/2}]^{1/2}
 \\
\label{eq:e26b}
k_2=(2m/\hbar^2)^{1/2}[E_F-(\varepsilon^2-\Delta^2)^{1/2}]^{1/2}
\end{eqnarray}
Since we are interested in excitations for which
$\varepsilon<\Delta$ , these wave vectors are complex and their
imaginary part ensures the exponential decay of the excitations
inside S. We can write them in the form:
\begin{eqnarray}
\label{eq:e27a} k_1=K_1+iK_2
 \\
\label{eq:e27b} k_2=K_1-iK_2
\end{eqnarray}
To a good approximation, $K_1 = k_F$ , and   $K_2 \ll K_1$ . It is
instructive to calculate explicitly the decay wave vector. To fix
the order of magnitude, we easily obtain it for $\varepsilon = 0$:
\begin{equation}
\label{eq:e28} K_2(\varepsilon=0) = \Delta/\hbar v_F
\end{equation}
Which, except for a factor of $\pi$ is the inverse of the
coherence length $\xi$. It is easily seen that the decay length
diverges when $\varepsilon$ approaches $\Delta$.

The ratios $(\beta_1/\alpha_1)$ and $(\beta_2/\alpha_2)$ are
determined by the BdG equations, so that there are 4 unknowns and
4 homogeneous linear equations relating them (the boundary
conditions at the interface). After some simplifications resulting
form the fact that $(\Delta/E_F) \ll 1$, the condition for the
existence of a solution to the eigenvalue problem reads:
\begin{equation}
\label{eq:e29} tan(k'_2 d_N)=tan(k'_1 d_N - \phi)
\end{equation}
where $\phi$ is defined by:
\begin{equation}
\Delta cos \phi = \varepsilon
\end{equation}
Taking into account Eq.\ref{eq:e24a} and \ref{eq:e24b}, solutions
of \ref{eq:e29} are those of:
\begin{equation}
(2d_N/\pi\xi)cos\phi=\phi+n\pi
\end{equation}
where we have used the relation $\xi = (\hbar v_F/\pi\Delta)$. For
any finite value of the normal slab thickness, there is no zero
energy solution. An energy gap has been induced in N by the
proximity with S. When the slab thickness is small compared to the
coherence length, there is only one solution at  $\varepsilon <
\Delta$ and it approaches $\Delta$. This bound state is in fact
localized over the length $\xi$, over which excitations can
penetrate inside S. When the thickness is much larger than $\xi,
\phi$,  approaches $\pi/2$. There is a large number of solutions,
the separation between the levels being:
\begin{equation}
\Delta\varepsilon=(\hbar^2/2m)(\pi k_F/d_N)
\end{equation}
As noted by Saint James \cite{Saint_James:1964}, this is half the spacing between
the electronic levels in a 1D isolated normal metal of the same
thickness. He explains this difference by noting that excitations
in N in contact with S have a mixed electron-hole character. At
energies smaller than $\Delta$, the coefficients $\alpha$ and an
$\beta$ are close to each other. Over a complete cycle comprising
two ASJ reflections at the N/S interface, and two specular
reflections at the free surface of N, the quasi-particle is
electron-like half of the time, and hole like the other half. This
same Saint James cycle leads, as we shall see, to the formation of
zero energy states at the surface of a d-wave superconductor.

In a more realistic 3D situation, instead of the discrete energy
levels that we have obtained, the DOS is finite at any finite
energy. Eigenvalues of the energy for quasi-particle trajectories
making an angle $\Theta$ with the normal to the interface are
reduced because the path traveled before the ASJ reflection takes
place is longer. These eigenvalues tend to zero when $\Theta$
approaches $\pi/2$, but the solid angle covered by such
trajectories tends itself to zero and as a result the DOS tends to
zero linearly with $\varepsilon$ (Fig.1).

\begin{figure}
\epsfxsize=3.5in \epsfbox{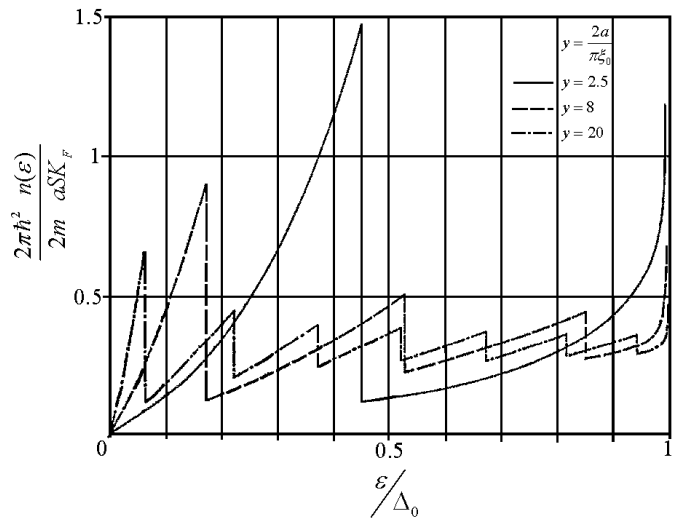}

\caption{Density of states for an N/S contact for different values
of the normal layer thickness a. The thickness is given in terms
of a normalized parameter ($2a/\pi\xi$). (after Saint James 1964
op.cit.).}
\end{figure}

\subsection{The case of a normal layer sandwiched between two superconductors. Effect of a phase difference.}
\label{sec:IIB}

It follows from our analysis that we would have obtained the same
eigenvalues if we had considered a normal slab of thickness $2d_N$
sandwiched between two superconductors. This is because in that
case one reflection at each interface is sufficient to complete a
cycle: an electron hitting the interface with $S_1$ is reflected
as a hole, which is reflected back as an electron by $S_2$. As a
matter of fact, this geometry was used to study the proximity
effect by thermal conductivity measurements of S/N/S sandwiches
\cite{Wolf:1971}. These measurements did reveal a reduced DOS in N.
But for what concerns us here, the main interest of an S/N/S
geometry is that it allows to look for the effect of a phase
difference between the two superconductors on the DOS. Such a
phase difference may for instance be induced by a current flowing
perpendicular to the interfaces. To be specific, let us consider
the case where this phase difference is equal to $\pi$ (this will
be the case of interest for a d-wave superconductor). The pair
potentials in $S_1$ and $S_2$ have then opposite signs. The BdG
equations impose that the ratios ($u/v$) will also have opposite
signs in $S_1$ and $S_2$. The continuity conditions at the
interfaces then require that solutions in N will be of the form:
\begin{eqnarray}
\label{eq:e213a} u = \alpha sin[k'_1 (x+d_N)]
 \\
\label{eq:e213b} v = \beta cos[k'_2 (x+d_N)]
\end{eqnarray}
The solution to the eigenvalue problem is then that of:
\begin{eqnarray}
\label{eq:e214} -cot(k'_2 d_N) = tan (k'_1 d_N - \phi)
\\
\label{eq:e215} (k'_2 d_N) - \pi/2 = (k'_1 d_N - \phi) + n\pi
\end{eqnarray}
which gives the result:
\begin{equation}
\label{eq:e216} (2d_N/\pi\xi)cos\phi = \pi/2 - \phi + n\pi
\end{equation}
Contrary to the case treated by Saint James, it is immediately
seen that here $\phi=\pi/2$ is a solution for any value of the
thickness. In other terms, there exists a solution with the
eigenvalue $\varepsilon = 0$ even in the limit where the thickness
of the normal slab is zero. In effect, one does not need a normal
layer to obtain a zero energy bound state when there is a change
of phase by $\pi$. This zero energy solution is localized near the
interface between the two superconductors, it decays in the
superconducting banks over a coherence length. It is a zero energy
interface bound state. A self consistent solution would of course
give a pair potential going to zero at the interface. The
situation is similar to that near a vortex core: on opposite sides
of the core, phase differ by $\pi$ , the pair potential goes to
zero at the center of the vortex core to accommodate this change
in the phase, and there are low lying states of extension $\xi$
\cite{Caroli:1964}.

\subsection{Surface bound states in a d-wave superconductor}
\label{sec:IIC}

The situation described in the previous sub-section is somewhat
artificial, particularly in the limit of a zero thickness normal
slab. But this exercise helps understand what happens at the
surface of a d-wave superconductor when it is oriented
perpendicular to a node direction. For the $d_{x^2 - y^2}$
symmetry, the pair potential is of the form:
\begin{equation}
\Delta=\Delta_0 cos2\theta
\end{equation}
where $\theta$ is the angle with one of the principal axis. The
pair potential is at a maximum along these axis, and changes signs
at 45 degrees from them. The pair potentials on either side of
these nodes have the same absolute values but opposite signs. Let
us go back to the original Saint James geometry and consider a
normal metal slab in contact with the surface of a d-wave
superconductor having the above orientation (Fig.2). An electron
in N moving towards the interface with a wave vector at some
finite angle with the interface will be ASJ reflected as a hole by
a pair potential having, say, the positive sign. This hole will
then be specularly reflected at the outer surface of N, after
which it will be ASJ reflected as an electron by a pair potential
having the negative sign. This electron will then in turn be
specularly reflected at the free surface of N, which will close
the Saint James cycle. It is equivalent to that treated in the
above sub-section: there are two successive ASJ reflections by
pair potentials having phases that differ by $\pi$. Zero energy
states are formed in N, and extend inside S over a coherence
length. This geometry was first studied by Hu \cite{Hu:1994} who showed
that zero energy states are formed even in the limit of a zero
thickness normal slab, as in the above exercise. In the
semi-classical approximation, they are zero energy surface bound
states, or ASJ bound states.

\begin{figure}
\epsfxsize=3.5in \epsfbox{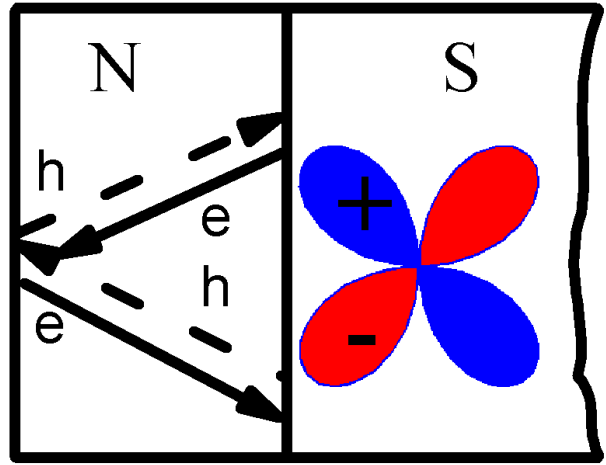}

\caption{Schematic representation of an ASJ cycle for a d-wave
superconductor coated with a normal metal layer, the interface
being oriented perpendicular to a nodal direction.}
\end{figure}

The spectroscopy study of these states will constitute a
substantial part of this review. Whenever detected they are a
clear signature of a pair potential that reverses sign around the
Fermi surface. They are modified by the presence of even a small
imaginary component. such as $is$ or $id_{xy}$, which cancels the
sign reversal in the vicinity of a d-wave node. In this way, such
components can be detected and their amplitude measured. In short,
the spectroscopy study of ASJ states allows the determination of
the detailed symmetry of the pair potential and that of the
respective amplitudes of its components.

It will be appreciated that ASJ bound states will be best studied
by tunneling from a normal metal electrode through a junction
formed directly at the surface of the d-wave superconductor having
the appropriate orientation. Making a clean contact with a normal
metal would result in short lived ASJ states.

ASJ spectroscopy of the HTSC thus employs two different contact
techniques: 1) clean Sharvin contacts are appropriate when formed
on surfaces oriented perpendicular to an axis along which the
order parameter is at its maximum, because in that geometry no ASJ
bound sates are formed and the amplitude of the gap is immediately
accessible; 2) tunneling contacts are preferred when formed on
surfaces oriented perpendicular to a (presumed) node direction, in
order to detect the presence of ASJ bound states due to an
unconventional order parameter symmetry. These two methods are
reviewed in the following sections.
\newpage
\section{CONDUCTANCE CHARACTERISTICS OF SHARVIN CONTACTS}
\label{sec:III}
\subsection{Sharvin contacts as a tool for ASJ spectroscopy}
\label{sec:IIIA}

Let us consider a clean contact between two metals having a very
small cross section $a^2$, so that its electrical conductance in
the normal state is equal to the number of quantum channels
connecting them, multiplied by the quantum conductance
$(e2/\hbar)$. Namely, we neglect for the time being any resistance
that might arise from a dielectric barrier between the two metals,
or from a mismatch of the Fermi velocities between them. Such a
situation can be nearly realized if we use broad conduction band
metals since they all have Fermi velocities of the order of
$1\cdot10^8$ cm/sec, and if we can avoid the formation of an oxide
at the interface. The current voltage relationship of this contact
is:
\begin{equation}
\label{eq:e31}
 I = (e^2/h)(k_Fa)^2 V
\end{equation}
The current density through the contact is:
\begin{equation}
J = nev
\end{equation}
where $n$ is the carrier density and v their velocity across the
contact. From \ref{eq:e31} and the current density definition $J =
(I/a^2)$, we obtain for the velocity the expression:
\begin{equation}
v = (eV/h) (k_F^2/n)
\end{equation}
When we use Sharvin contacts to perform ASJ spectroscopy, the bias
across the contact will reach values of the order of the pair
potential. Using the value for the carrier density in the free
electron model, we obtain at such bias a velocity of the order of:
\begin{equation}
v = (\Delta/p_F)
\end{equation}

which is the depairing velocity in the superconducting side with
gap $\Delta$.

This high velocity is the main reason why we must use a contact
size smaller than the electron mean free path in order to avoid
heating effects, and also a contact size smaller than the
coherence length (Fig.3). Spreading of the current after the
contact bottleneck then reduces the current density below the
depairing value already at distances smaller than $\xi$, thus
avoiding quenching superconductivity at the contact, since
superconductivity cannot be quenched over a length scale smaller
than $\xi$.

\begin{figure}
\epsfxsize=3.5in \epsfbox{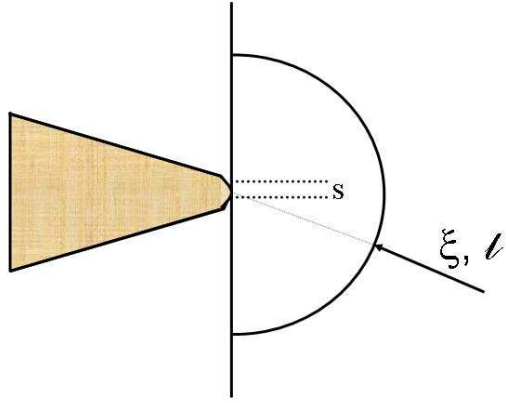}

\caption{Schematic representation of a Sharvin Point Contact
having a size s much smaller than the coherence length $\xi$ and
the mean free path $l$. At biases of the order of the gap, the
current density at the contact is of the order of the depairing
value, but at distances of the order of ($\xi,l$) it is reduced
much below that value. The condition $s \ll (\xi,l)$ avoids
heating effects and quenching of superconductivity in the vicinity
of the contact.}
\end{figure}

In summary, a small contact size is favorable for three reasons:
i) the condition $a\ll l$  makes the contact ballistic, which
prevents heating effects at the large current densities reached at
biases of the order of the gap; ii) the condition $a < \xi$ plays
two roles: it prevents the weakening of superconductivity in S due
to the proximity with N; and it prevents the destruction of
superconductivity at biases of the order of the gap, where the
carrier velocity reaches the depairing value $(\Delta/p_F)$.

\subsection{Fabrication of Sharvin contacts}
\label{sec:IIIB}
In a clean superconductor the mean free path $l >
\xi$, and the above conditions are met if $a < \xi$. In a LTSC,
$\xi$ is typically of the order of $1000\AA$, and the contact
qualifies as a Sharvin contact if a $\approx 100\AA$. For a HTSC,
we would rather require a $=approx 10\AA$. When making a Point
Contact, how can we make sure that such conditions are met?

The fabrication of point contacts has been well described in the
literature, see for instance BTK \cite{Blonder:1982}. In short, a relatively
sharp metallic tip, having a local radius of curvature of the
order of 1 micron, is brought delicately in contact by a
mechanical device with a bulk counter-electrode. The tip can be
for instance made of a thin Gold wire , cut with a sharp razor
blade. If the actual size of the contact were of the order of the
radius of curvature of the tip, the above conditions would not be
met. Its electrical resistance would be of the order of
1m$\Omega$. In fact, the resistance of the contact is often found
to be of the order of 10 to 100$\Omega$. But how can we determine
whether this larger value reflects actually a small contact size,
or a dirty contact? We can get an answer to this question if we
combine the value of the normal state resistance $R_N$  (namely,
its value above the critical temperature, or more practically its
value well above the gap bias) with its value $R_S$ at low bias
(below the gap). In an ideal contact, we would have $(R_N/R_S) =
2$. This is of course never achieved. But a ratio $(R_N/R_S) > 1$,
is an indication that ASJ reflections may dominate,  although some
junction structures (such as a proximity effect across the
junction) may give such ratios and should be carefully checked for
instance through the bias dependence of the conductance.

It turns out that in many cases the contact realized is indeed
clean, and that its size falls in the range of a few $10\AA$. One
may wonder how that can happen, in view of the rather crude
contact technique used here. The practitioners know from
experience that when the tip is first brought in contact with the
counter-electrode, the resistance is usually fairly high, in the
range of a few k$\Omega$, and the I(V) characteristic is
structureless. By applying some slight movements to the tip, it is
however possible to bring the resistance down to the interesting
range of 10 to 100$\Omega$ , and to obtain meaningful
characteristics. One may conjecture that these movements scratch
away some of the insulating material at the surface, revealing the
underlying pristine material. This, however, does not explain the
very small size of the contact achieved. On the other hand, it is
well known that when trying to produce a uniform tunneling
barrier, one often encounters a problem of shorts presumably due
to pinholes in the barrier. It may be that small good contacts are
established through some naturally occurring defects in the
insulating layer, but this is still imperfectly understood. Small
size contacts have been observed using very different techniques
of tip preparation, such as electro-chemical etching of a Nb wire,
or cutting a thin Au wire with a sharp razor blade, as said
before. A review of these methods of tip preparation can be found
in Achsaf \emph{et al.}, 1996.
\subsection{BTK model}
\label{sec:IIIC}
An electron moving from the N side towards the
interface can be scattered in 4 different ways:

i)  it can be reflected as a hole along the incident trajectory
(ASJ reflection) with probability $A(\varepsilon)$.

ii) it can be reflected as an electron (normal specular
reflection) with probability $B(\varepsilon)$.

iii)    it can be transmitted as an electron having a momentum $k
> k_F$ (no branch crossing) with probability $C(\varepsilon)$.

 iv) it can be transmitted as an electron having a momentum $k < k_F$ (branch
crossing) with probability $D(\varepsilon)$. The sum of these
probabilities must be equal to one:
\begin{equation}
\label{eq:e35}
A(\varepsilon) + B(\varepsilon) +C(\varepsilon)
+D(\varepsilon) = 1
\end{equation}
Since current is conserved across the interface, it suffices to
calculate it for instance at the N side of the contact:
\begin{equation}
\label{eq:e36}
 I=J_0\int_{-\infty}^{+\infty}
[1+A(\varepsilon)-B(\varepsilon)][f(\varepsilon-eV)-f(\varepsilon)
d\varepsilon
\end{equation}

where $f(\varepsilon)$ is the Fermi function and $J_o$ is a
conductance taking into account the geometry of the contact.

BTK have given a complete calculation of the dependence of
coefficients $A$ and $B$ on energy for barriers characterized by a
$delta$ function potential. For their derivation an we refer the
reader to their paper. Here below we limit ourselves to some
simple limiting behaviors which we feel are of particular
importance.

\subsubsection{Case of a clean interface}
For a clean interface, there are no specular reflections at the
interface, $B(\varepsilon)=0$. Also, $A(\varepsilon)=1$ for
$\varepsilon < \Delta$. For $\varepsilon > \Delta$, excitations
can propagate in S. They have a partial electron character with
amplitude $u_o$, and a partial hole character with amplitude
$v_o$. From BCS theory, we know that:
\begin{eqnarray}
\label{eq:e37a} u_0^2
=[1+(\varepsilon^2-\Delta^2)^{1/2}/\varepsilon]/2
 \\
\label{eq:e37b} v_0^2 =
[1-(\varepsilon^2-\Delta^2)^{1/2}/\varepsilon]/2
\end{eqnarray}

ASJ reflections occur in the proportion of the probabilities for
hole to electron characters of the excitations propagating in S:
\begin{equation}
\label{eq:e38}
A(\varepsilon)=[1-(\varepsilon^2-\Delta^2)^{1/2}/\varepsilon]/[1+(\varepsilon^2-\Delta^2)^{1/2}/\varepsilon]
\end{equation}
The conductance is equal to twice the normal state value below the
gap, and goes back to it over a scale $\Delta$ in a manner that
can be calculated from Eq.\ref{eq:e38} and Eq. \ref{eq:e36}.

\begin{figure}
\epsfxsize=3.5in \epsfbox{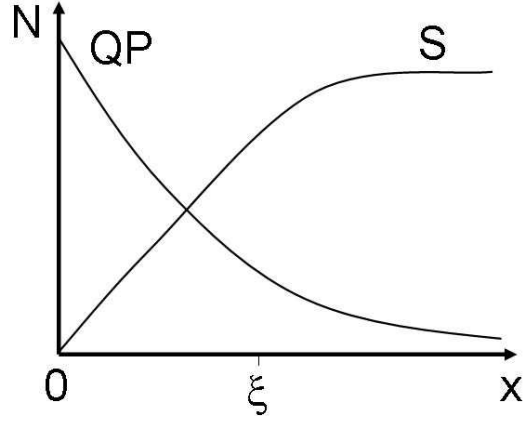}

\caption{Build up of the superfluid density S and decrease of the
quasiparticle density QP in the S side of and N/S contact. ASJ
reflections build up progressively over the distance $\xi$.}
\end{figure}

It is interesting to consider more closely how the normal current
in N is converted into a superfluid pair current in S (Fig.4). As
already shown by Saint James, and noted earlier in this review,
there are evanescent quasi-particle waves in S at excitation
energies smaller than the gap. They decay over a length scale of
the order of $\xi$, and are at the same time converted into
superfluid. More precisely, BTK show that the decay length is
given by:
\begin{equation}
\lambda=(\hbar v_F/2\Delta)[1-(\varepsilon/\Delta)^2]^{-1/2}
\end{equation}
This progressive conversion of quasi-particles into superfluid has
important consequences, some of them we have already outlined. At
a distance $\xi$ from the interface, the current density is
reduced by a factor $(a/\xi)^3$, which in a LTSC can be a factor
of $1\cdot10^{-6}$. The velocity is then negligible compared to
the depairing velocity even at biases much larger than the gap.
The situation is evidently much less favorable in the HTSC because
of their short coherence length. Then, we can hope at small bias
to have at best $a\leq\xi$. However, as the bias is increased the
situation becomes more favorable because of the divergence of the
decay length. Another point, which is trivial for LTSC, is that
propagation of quasi-particles over a distance $\xi$ from the
interface is necessary for the conversion to take place, and
therefore for ASJ full reflections to occur. This condition is not
a trivial one when we consider some situations peculiar to the
HTSC, such as the existence of a pseudogap that may be larger than
the pair potential. This special situation will be discussed in a
later section.

\subsubsection{Contacts with a finite transparency}
BTK have calculated the coefficients $A(\varepsilon)$ and
$B(\varepsilon)$ for contacts with a finite transparency which
they have modeled with a $\delta$ function barrier $V =
H\delta(x)$. They have used in their calculation a dimensionless
parameter $Z = (H/\hbar v_F)$. For the case where the Fermi
velocities in N and S are different, Blondeer and Tinkham \cite{Blonder:1983}
have shown that one can replace $Z$ by an effective barrier
parameter:
\begin {equation}
Z_{eff}=Z^2 + (1-r)^2/4r^2
\end{equation}
Where $r = v_{FN}/v_{FS}$ (or the inverse). The shape of the I(V)
characteristics is a function of $Z_{eff}$ only. There is no way
one can distinguish between the effects of a dielectric barrier
and that of a mismatch between the Fermi velocities. This result,
obtained for a  function barrier, is not general. In particular,
it does not hold when the normal side is spin polarized (see the
last section of this review).

A finite $Z$ results in a finite probability of specular electron
reflections at the N/S interface. $B(\varepsilon)$ is now finite,
$A(\varepsilon) < 1$ even at $\varepsilon < \Delta$: the
conductance below the gap is smaller than $2R_N^{-1}$, it goes to
zero as $Z$ is made very large. BTK have shown that at zero-bias:
\begin{equation}
A(0)=(1+2Z_{eff}^2)^{-2}
\end{equation}
Since at biases smaller than the gap $C = D = 0$, it follows from
the sum rule $A +B +C +D = 1$ and Eq.\ref{eq:e36} that the current
at zero-bias is proportional to $2A(0)$. In the normal state, or
at bias much larger than the gap, it is proportional to $(1 - B) =
(1 +Z_{eff}^2)^{-1}$. From Eq.\ref{eq:e35} the ratio of the
zero-bias to the high bias resistances, $(R_S/R_N)$ is thus given
by:
\begin{equation}
R_S/R_N)=2(1+Z_{eff}^2)/(1+2Z_{eff}^2)^2
\end{equation}
From the measurement of the ratio of the contact resistances at
high and zero-bias one can calculate the value of the effective
barrier parameter. This value is useful for two purposes. First,
it allows the calculatation of the actual size of the contact by:
\begin{equation}
(k_Fa)^2 = (h/R_Ne^2 )(1 + Z_{eff}^2)
\end{equation}
Second, this parameter gives a lower bound to the ratio of the
Fermi velocities by assuming the absence of any dielectric barrier
at the interface.

\begin{figure}
\epsfxsize=3.5in \epsfbox{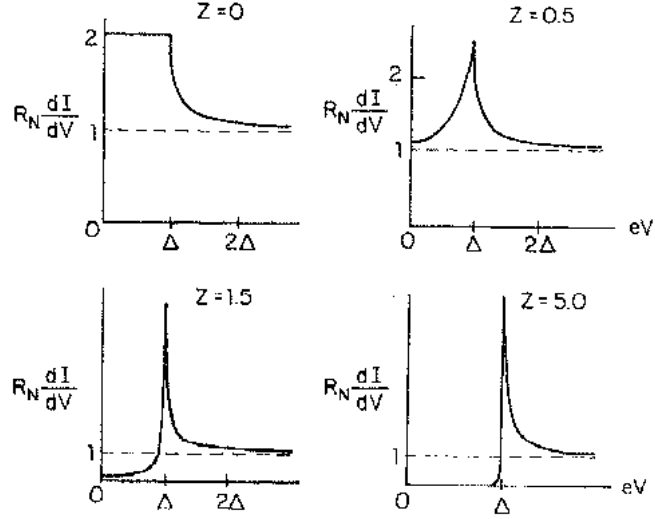}

\caption{Conductance characteristics of N/S contacts for various
values of the barrier parameter $Z$ (after
\citealt{Blonder:1982}).}
\end{figure}

More generally, using the analytical expressions derived by BTK
for the coefficients $A(\varepsilon)$ and $B(\varepsilon)$, it is
possible to fit experimental I(V) curves and to extract values of
the gap and of the barrier parameter. Fig.5 gives a few examples
of conductance characteristics calculated for different values of
the barrier parameter. As it increases, the normalized zero-bias
conductance falls below 2, while at the gap edge it increases
above 2. In fact, at the gap edge the conductance value is
unaffected by the barrier. BTK have obtained values of the gap and
of the $Z$ parameter by fitting the conductance curves of Nb/Cu
Sharvin contacts to the BTK theory \cite{Blonder:1983}.
Characteristics could be fitted successfully for normal state
resistances falling in the range of 10 to 100$\Omega$. Reported
values of $Z_{eff}$ are smaller than 1, and can be as small as
0.3. Calculated contact radii vary from 10 to 120$\AA$, fully
qualifying them as Sharvin contacts meeting the conditions $a <
(\xi, l)$. The bound to the mismatch of the Fermi velocities is as
expected. In fact, the experimentally determined values of the
effective barrier parameter mean that the contact is basically a
clean one. In turns, this justifies modeling the barrier as a
$\delta$ function, since the mismatch of the Fermi velocities
occurs over an atomic distance.

\subsection{HTSC Point Contact results for an anti-nodal
orientation.}
\label{sec:IIID}
\subsubsection{Early experiments on YBCO single crystal quality
samples} The earlier Sharvin point-contact experiments on single
crystal quality samples were performed  by Hass \emph{et
al.}\cite{Hass:1992} on melt textured YBCO samples cut out in a cubic shape
so that four faces had the  (100) or equivalent orientation, and
two the (001) orientation. As already noted, the main motivation
for attempting this experiment was the hope that there was a
better chance to obtain a good point contact than to make a good
tunnel junction. They revealed the following main features:

-   on (100) faces, contacts with a normal state resistance of
about 10$\Omega$ could be made. Their conductance increased by
about 50\% below a bias of about 20 mV (Fig.6 ). This bias value
was interpreted as being the gap edge. The shape of the
characteristic was generally in accordance with the predictions of
BTK for a barrier parameter of about 0.3, except for two features.
First, the return to the normal state conductance above the gap
was somewhat faster than it should have been; this might have been
due to the fact that the condition $a <\xi$ was only barely met,
because of the short coherence length. Second, the data did not
show the expected conductance peak at the gap edge. The absence of
this conductance peak was also noted later on similar contacts
produced on LSCO samples \cite{Hass:1994}. The main
surprise came from the substantial enhancement of the conductance
at low bias, which implied a small mismatch of the Fermi
velocities.

\begin{figure}
\epsfxsize=3.5in \epsfbox{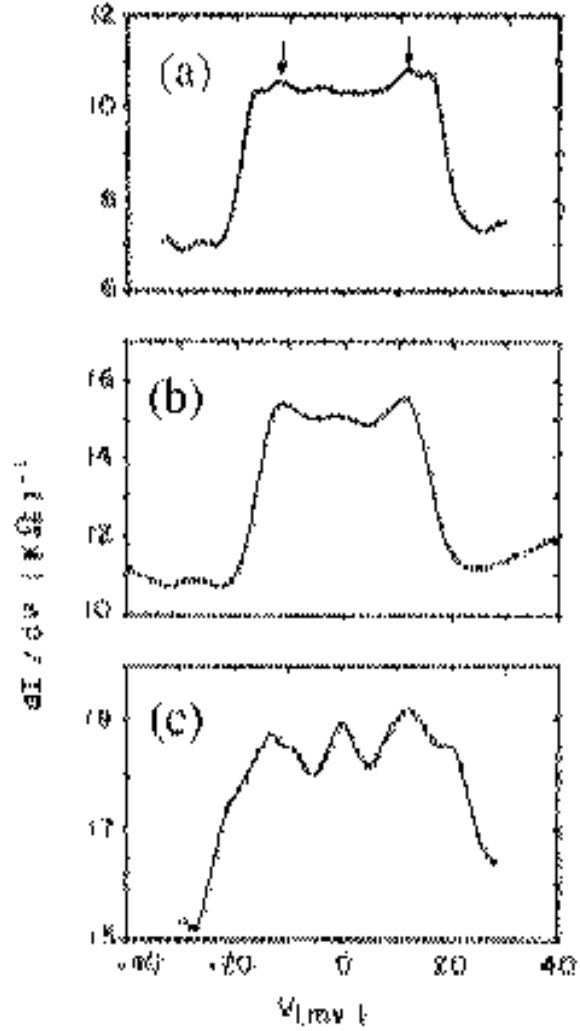}

\caption{Conductance characteristics of Au/YBCO Point Contacts for
the anti-nodal orientation (after \citealt{Hass:1992}). The
increase of the conductance at low bias is of about 50\%. The edge
is at about 20 mV. Note the absence of a conductance peak at the
gap edge. Data from three different locations on the same sample.}
\end{figure}

- on (001) oriented faces, the characteristics were basically
structureless except for an occasional zero-bias peak, and
positive slopes at negative as well as positive biases. The
absence of a conductance enhancement at low bias for this
orientation was expected, in view of the large mismatch of the
Fermi velocities for that case. However, BTK theory then predicts
that the characterisitic should have the shape of a regular
Giaever tunneling junction, which was not observed.

\subsubsection{Effect of the d-wave symmetry for an anti-nodal
direction}
 The original theory of BTK is not appropriate for
d-wave symmetry. However, the later does not have dramatic effects
for an anti-nodal direction, probed in the experiments of Hass
\emph{et al.} \cite{Hass:1992}. As noted above, there are no ASJ low energy
surface bound states for that orientation, because specular
reflections at the surface then preserve the value and the sign of
the pair potential. The coefficients $A$ and $B$ can be calculated
by performing a 2D integral of the BTK coefficients over all
angles. For clean contacts, having an effective $Z$ parameter
smaller than 1, it is not necessary to take into account a finite
tunneling cone aperture. Typical conductance characteristics
shapes are shown Fig.(7) ($Z$ values: 0; 0.2; 0.3; 0.5; 0.7; 1.0).
For $Z = 0$, they assume a triangular shape; for $Z$ values around
0.3, they are rather flat up to the gap edge, this is the case
observed in the early experiments on YBCO shown above; for $Z$
values in the range of 0.5 to 1, they assume a V shape at low
bias, reaching a maximum slightly below the gap, followed by a
sharp descent back to the normal state conductance. It is
noticeable that for an anti-nodal direction, in this range of $Z$
values which is typical for Point Contacts, the maximum
conductance reached is always smaller than twice the normal state
value. Typical maximum conductance values are around 30 to 50\%
higher than the normal state value.

\begin{figure}
\epsfxsize=3.5in \epsfbox{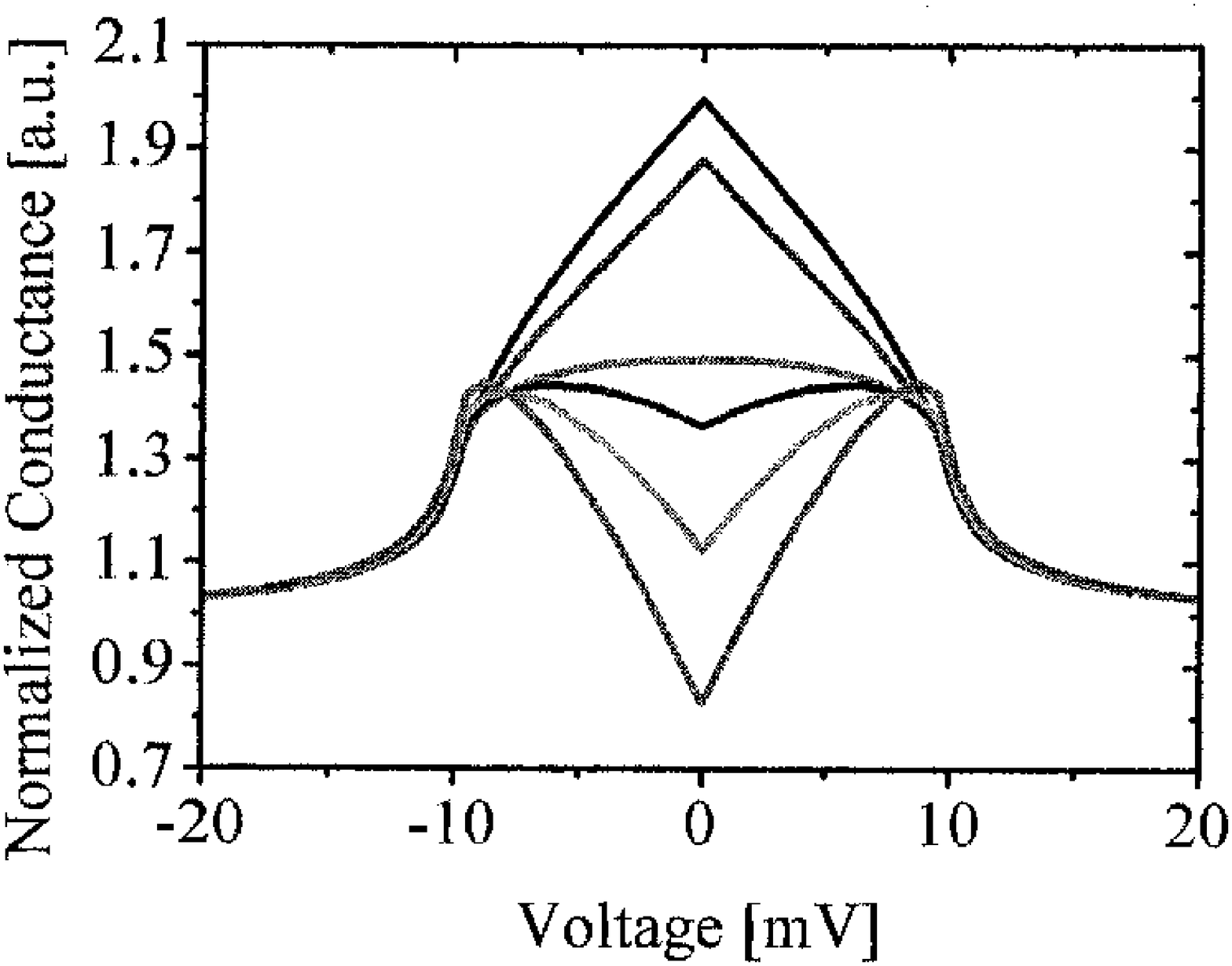}

\caption{Calculated conductance characteristics for a d-wave order
parameter in an anti-nodal orientation, at different values of the
barrier parameter $Z$ (0; 0.2; 0.3; 0.5; 0.7; 1.0). Note for $Z >
0.5$ the clear V-shape at low bias and the sharp decrease of the
conductance at the gap edge.}
\end{figure}

The d-wave symmetry explains the main disagreement between the
data of Hass \emph{\emph{et al.}}\cite{Hass:1992} and the original
BTK theory, namely a conductance at the gap edge smaller than
twice the normal state value. Actually, the various theoretical
shapes calculated for the range $0 <Z <1$ have been observed
experimentally. Fig.(8) shows data obtained on an a-axis oriented
YBCO surface \cite{Kohen:2003} and Fig.(9) data obtained on a
BiSrCaCuO single crystal in the (100) orientation
\cite{DGorno:1998}, with fits to theory. The predicted V-shape at
low bias is clearly observed in both cases. Fits are of a high
quality, and require only a small "smearing" factor $\Gamma$
\cite{Kohen:2003}. They establish that ASJ spectroscopy is a
reliable and quantitative spectroscopic tool for the study of
HTSCs.  When performed in an anti-nodal direction, it provides a
precise determination of the gap value. The effect of the d-wave
symmetry is also clearly seen in the shape of the characteristics.

\begin{figure}
\epsfxsize=3.5in \epsfbox{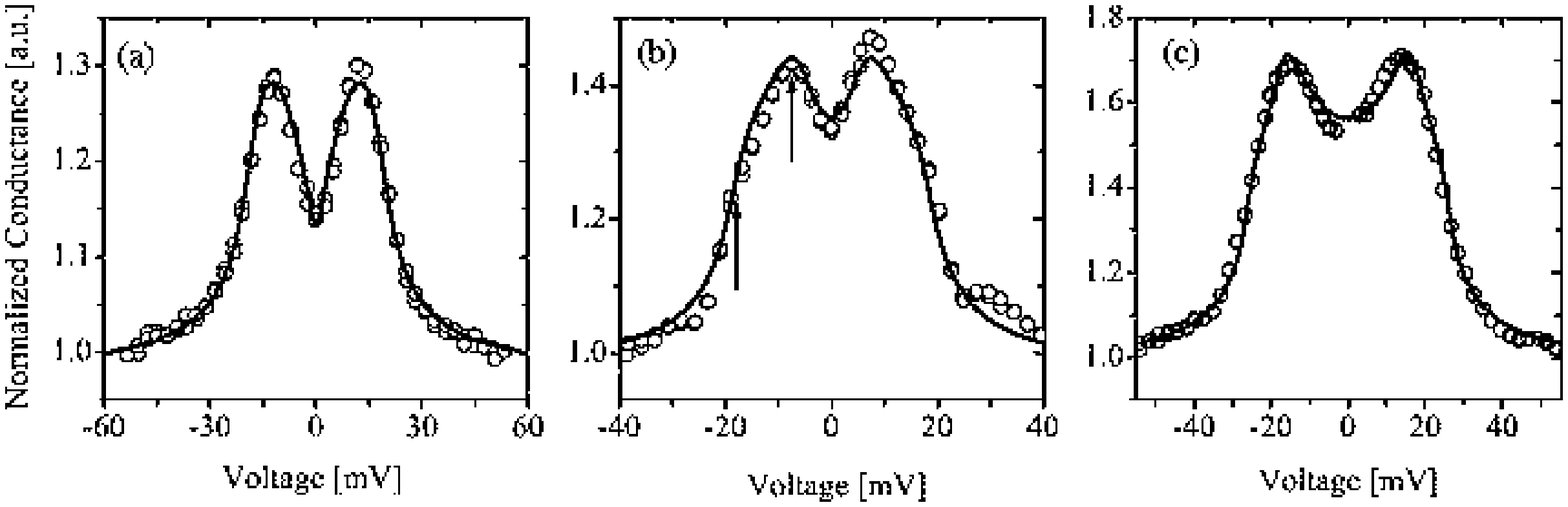}

\caption{Conductance characteristics of Au/YBCO films contacts
fitted to a d-wave order parameter for different values of the
barrier parameter $Z$ (from left to right: 0.68; 0.49; 0.34). The
fit is for an anti-nodal direction. Note the well pronounced
V-shape at low bias for the highest $Z$ contact. (after
\citealt{Kohen:2003}).}
\end{figure}

\begin{figure}
\epsfxsize=3.5in \epsfbox{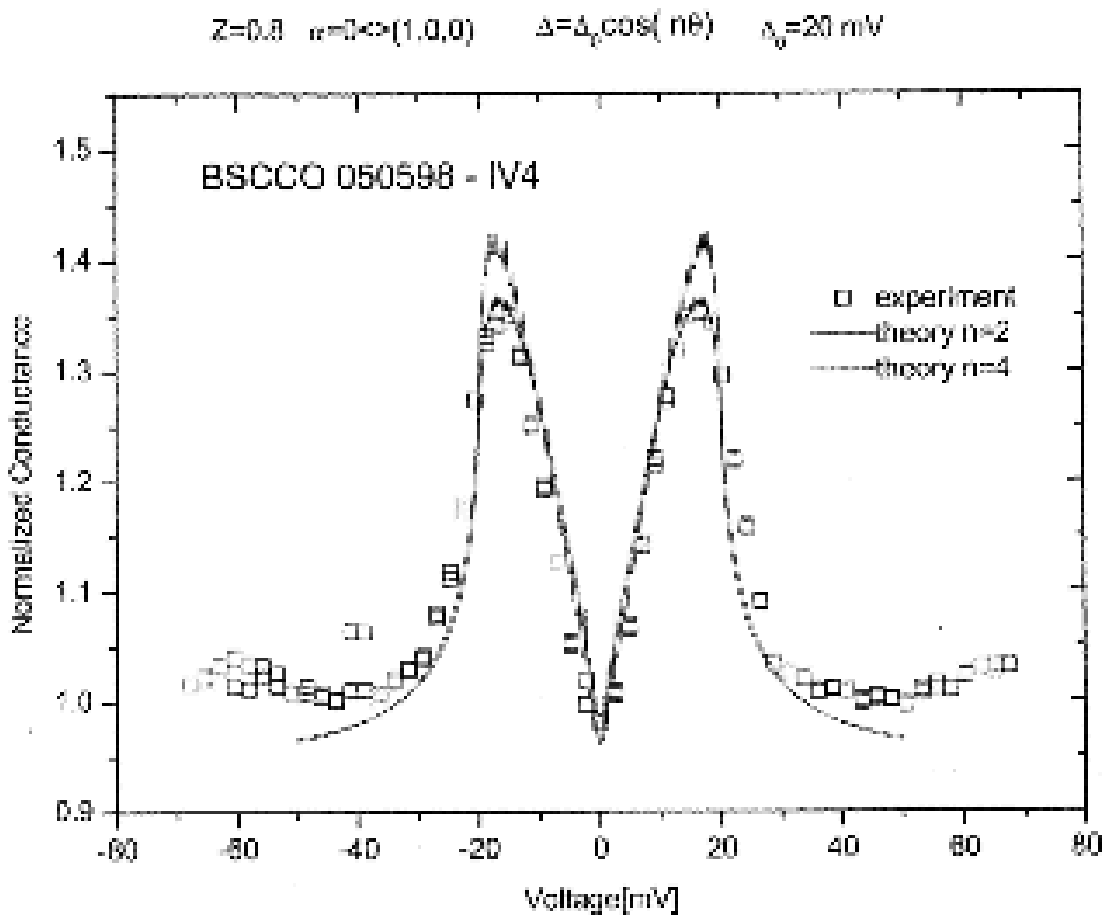}

\caption{Measured conductance characteristic of a
$Au/Bi_2Sr_2CaCu_2O_8$ in-plane contact to a bulk oriented sample.
Note the low bias V-shape and the sharp drop at about 20 mV. The
fit is for an anti-nodal direction and $Z = 0.8$. (After I. Okashi
and A. Kohen privet communication)}
\end{figure}

Two kinds of deviation from pure d-wave behavior have been
reported. A special proximity effect has been observed between the
normal tip and the d-wave superconductor at very small $Z$ values
($Z < 0.5$), which may induce an is component in the
superconductor. This is discussed in detail in the last section of
this review. Another deviation has been seen in strongly overdoped
YBCO samples. It involves a small imaginary minority component
($id_{xy}$ or $is$), whose occurrence is also discussed in the
last section.

\subsection{Renormalization of the Fermi velocity}
\label{sec:IIIE}
\subsubsection{The problem of the small Fermi velocity mismatch}
The other question raised by the experiments of Hass \emph{et al.}
was the surprisingly small value of the barrier parameter
needed to fit some of the data. $Z$ values of 0.3 to 0.4, as found
for YBCO \cite{Hass:1992,Wei:1998} imply that the ratio between the Fermi velocities of YBCO
and of the normal tip is at most a factor of 2 (assuming that the
finite $Z$ is entirely due to the Fermi velocity mismatch). Taking
for instance the case of the Au tip, where the Fermi velocity is
$1.4\cdot10^8$ cm/sec, one obtains for the Fermi velocity in YBCO
a lower bound of $0.7\cdot10^7$ cm/sec. This is more than three
times larger than the velocity measured by ARPES in BiSrCaCuO,
$v_{FS} = 2\cdot10^7$ cm/sec (see for instance \cite{Margaritondo:1998}),
which is presumably typical of all HTSC. Using this value of
$v_{FS}$, one obtains $Z = 6$. For such a high $Z$ value, the
point-contact characteristic should be a tunneling one \cite{Tanaka:1996}, namely the small bias conductance should be much smaller
than the normal state value, contrary to what is observed
experimentally.

\subsubsection{Solution to the problem}
As shown by Deutscher and Nozieres \cite{Deutscher:1994}, the solution to the
problem lies in the fact that the small value of the
quasi-particle Fermi velocity in the cuprates is a many body
effect, which does not come into play in the mismatch that governs
ASJ reflections. The quasi-particle velocity is:
\begin{equation}
v_F = z \overline{v}_F
\end{equation}
 where:
\begin{equation}
\overline{v}_F = v_{F0} - \partial\Sigma/\partial k
\end{equation}
and:
\begin{equation}
\label{eq:e316}
z = 1/ (1 + \partial\Sigma/\partial\omega)
\end{equation}

$v_{F0}$ is the bare Fermi velocity (the band velocity, undressed
for interaction effects), $\Sigma$ is the self-energy correction
$\Sigma(k,w)$. The wave vector dependence of the self energy is a
non-local effect (usually quite small in metals), and its energy
dependence is a retardation effect, leading to mass enhancement.
This factor can sometimes be extremely large, like in heavy
fermions \cite{Hasselbach:1993}. It is in principle
accessible by a measurement of the low temperature electronic heat
capacity. In practice, however, that is not possible in the
high-$T_c$ cuprates, because of their extremely high critical
field. Another quantity that is sensitive to the mass enhancement
factor is the coherence length, since it is the quasi-particle
velocity that enters into that length:
\begin{equation}
\xi=\hbar v_F/\pi\Delta
\end{equation}
where $\Delta$ is the measured gap. We can obtain the value of
$v_F$ from Eq.\ref{eq:e316}, putting in the value of $\xi$ derived
from the measured upper critical field $Hc_2 = \phi_o/2\pi\xi^2$,
and the value of $\Delta$ obtained for instance from ASJ
spectroscopy, as described in the previous paragraph. We can then
calculate the mass enhancement factor $z$ by comparing this value
of $v_F$ to the lower bound of the velocity obtained from the
effective barrier parameter $Z_{eff}$ determined from a fit to
point-contact conductance characteristics.

If we take the specific example of YBCO, from $\xi = 15\AA, \Delta
= 20$ meV, we calculate for the quasi-particle velocity $v_F = 1.5
10^7$ cm/sec. The lower bound of the velocity obtained from the
lowest measured $Z_{eff} = 0.3$ is, from Hass \emph{et al.} 1992,
6 to $7\cdot10^7$ cm/sec, giving a mass enhancement factor of 4 to
5. It should be emphasized that this factor combines two effects:
non-locality and retardation. The respective contributions of
these two effects cannot be determined directly. This would
require a fairly exact knowledge of the bare band velocity. From
Massida \emph{et al.} \cite{Massida:1991} it is smaller than the velocity
derived from the point-contact measurements, suggesting that both
contributions are important.

\subsection{Concluding remarks}
\label{sec:IIIF}

In this Section, we have reviewed the application of the BTK model
to the study of ASJ reflections. We have shown that it can be
applied successfully, in a quantitative way, to point-contact
experiments carried on HTSC in configurations where phase effects
due to the d-wave symmetry of the order parameter are not
dominant. The BTK model is based on a solution of the Bogoliubov -
de Gennes equations for the pair potential. Use of these equations
assumes implicitly a Fermi liquid description of the HTSC. The
high quality of the fits between experiment and theory shows that
this assumption is justified, at least for samples near optimum
doping. The exact shape of the point-contact conductance
characteristics obtained on surfaces perpendicular to an
anti-nodal direction is in excellent agreement with a d-wave order
parameter. The small value of the barrier parameter $Z$ implies a
good match between the Fermi velocities of the normal metal tip
and the HTSC, which is well explained by a mass enhancement effect
in the spirit of Fermi liquid theory.

Geometries where the d-wave symmetry has a more dramatic effect
are reviewed in the next section.
\newpage
\section{ASJ SURFACE BOUND STATES}
As briefly introduced in \ref{sec:IIC}, zero energy ASJ surface
bound states are a direct result of a d-wave symmetry of the order
parameter when the surface is oriented perpendicular to a nodal
direction ( (1,1,0) oriented surface). The origin of these states
lies in the sign reversal of the pair potential "seen" by
quasi-particles upon specular reflection at the surface. For
reasons of symmetry, sign reversal for this orientation will occur
for trajectories making any angle with the normal to the surface.
For other surface orientations, sign reversal will occur for a
certain angular range, with the exception of the case of the
anti-nodal orientation, reviewed in IIID, where there is no sign
reversal for any trajectory. Hence, zero energy surface bound
states will exist for any surface orientation, except for the
anti-nodal one. In the (more practical) case of diffuse
reflections at the surface, for any orientation of the surface
there will always be a sign reversal for some trajectories.
Surface bound states are therefore a robust property of d-wave
superconductivity.
\subsection{Zero Bias Conductance Peak and d-wave symmetry}
\label{sec:IVA} Motivated by the findings of Hu \cite{Hu:1994}, Kashiwaya
and Tanaka (KT) \cite{Kashiwaya:1995} have extended the model of BTK to the case
of a d-wave symmetry, for all surface orientations.

The main difference with the results of BTK, obtained for the
s-wave symmetry case, comes about when one considers the two
transmission channels having the respective transmissions
probabilities $C(\varepsilon)$, for an electron-like transmission,
and $D(\varepsilon)$ for a hole like transmission. While in the
s-wave case a 1D calculation was sufficient, automatically taking
care of momentum conservation in the direction parallel to the
interface, here a 2D calculation is necessary.
\subsubsection{KT results for the (110) orientation}
We summarize here KT's results for the nodal orientation. We
follow KT's notation and define a normalized conductance:
\begin{equation}
\sigma(\varepsilon)=\overline{\sigma}_S(\varepsilon)/\overline{\sigma}_N(\varepsilon)
\end{equation}
where:
\begin{equation}
\overline{\sigma}_i(\varepsilon)= \int_{-\pi/2}^{+\pi/2}
\overline{\sigma}_i(\varepsilon,\phi)d\phi, (i=N,S)
\end{equation}
$\phi$ being the angle with the normal to the surface.

In the limit $Z \gg 1$, KT's expressions reduce to:
\begin{equation}\label{eq:e43}
\overline{\sigma}_N(\varepsilon,\phi)=4cos^2\phi/Z^2
\end{equation}
\begin{equation}\label{eq:e44}
\overline{\sigma}_S(\varepsilon,\phi)=32cos^4\phi/|4cos^2\phi+Z^2(1+\Gamma^2)|^2
\end{equation}
with:
\begin{equation}\label{eq:e45}
\Gamma=\varepsilon/|\Delta|-[(\varepsilon/\Delta)^2-1]^{1/2}
\end{equation}
At zero-bias, we have just $\sigma_S(0,\phi) = 2$. The conductance
is twice as large as the value it would have \emph{in the normal
state in the absence of any barrier} \cite{Deutscher:1998}.
On the other hand, the actual normal state conductance varies as
$Z^{-2}$. Hence the appearance in the measured conductance of what
is called a Zero Bias Conductance Peak, or ZBCP, for a high
barrier contact. The bias range over which the conductance is
substantially enhanced compared to its actual normal state value
is:
\begin{equation}\label{eq:e46}
|\varepsilon/\Delta|\leq Z^{-2}
\end{equation}
As the barrier height is increased, the conductance peak becomes
higher and narrower. Life time effects and discrete lattice
effects on the ZBCP have been discussed by Walker and Pairor
\cite{Walker:1999}. They broaden and reduce the ZBCP. Discrete lattice effects
limit zero energy bound states to certain orientations of the wave
vector. They can also modify the shape of the ZBCP near zero-bias.
There can thus be substantial deviations from the KT expressions.
Since the conductance at zero-bias $G_S(0)$  is unaffected by the
presence of a barrier for a pure (110) orientation, its value may
be used to calculate the actual size of the contact:
\begin{equation}\label{eq:e47}
(k_Fa)^2=(\hbar/e^2)G_S(0)
\end{equation}
Fig.(10) shows how the shape of the conductance characteristic
evolves as a function of $Z$ for the (110) orientation. For $Z =
0$, it has a triangular form, the conductance at zero-bias being
twice as large as the normal state value, reached when the applied
bias is equal to the value of the gap. For $Z \gg 1$,  the
conductance dips below its normal state value before returning to
it at a bias of the order of the gap. Notice that the gap is not
marked by a sharp structure at any value of $Z$.  Hence, the (110)
orientation is not as favorable as the (100) one for an accurate
determination of the gap. On the other hand, it is highly
sensitive to the symmetry of the order parameter.

\begin{figure}
\epsfxsize=3.5in \epsfbox{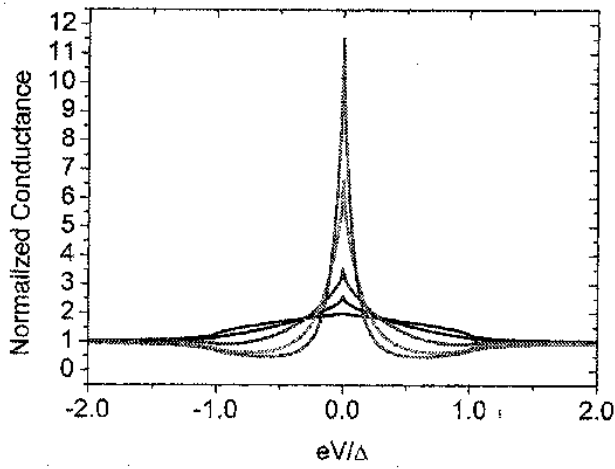}

\caption{Calculated conductance characteristics for a contact to a
d-wave superconductor in a nodal direction, for different values
of the barrier parameter $Z$ (0; 0.5; 1.0; 2.0; 3.0). Contrary to
the case of anti-nodal contacts, there is no sharp structure at
the gap bias, only a smooth return to the normal state
conductance.}
\end{figure}

\subsubsection{KT results for arbitrary orientations}
KT \cite{Kashiwaya:1995} have given expressions for the amplitudes of holes
$a(\varepsilon,\phi)$ and electrons $b(\varepsilon,\phi)$
reflections. These expressions allow to calculate the I(V)
characteristics for any surface orientation and $Z$ value, taking
into account a possible angular dependence of $Z$ in the case of a
strong barrier (tunneling cone).  The ZBCP is a robust feature of
the d-wave symmetry (Fig.11) \cite{Yang:1994}. It occurs for
any orientation of the surface, except the (100) one, and for any
value of $Z$. The structure at the gap edge is in general a weak
one. It can be a small step down at low $Z$ values, or a small
step up at large $Z$ values, or a weak maximum at large $Z$ values
and intermediate orientations. An important difference with a
contact to an s-wave superconductor is that no large conductance
peak is predicted at the gap edge (even for the (100) orientation
it remains of modest height). This peak, called the coherence peak
in s-wave superconductors, is destroyed for in-plane tunneling by
the very interference effects that give rise to the ZBCP.

\begin{figure}
\epsfxsize=3.5in \epsfbox{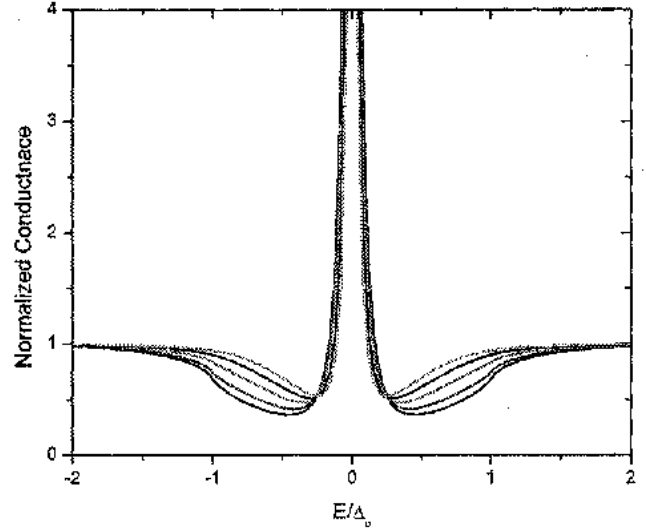}

\caption{Calculated conductance characteristics for a contact to a
d-wave superconductor in a nodal direction with $Z$ = 5, at
different openings of the tunneling cone. Return to the normal
state conductance always occurs at about the gap value. The curves
have been calculated using the weight function
$exp[-(\theta/\theta_M)^2 ]$, with $\theta_M$ values 90, 57,
33,23,18 degrees. }
\end{figure}

\subsection{Experimental results} \label{sec:IVB}
We review successively results obtained for low $Z$ contacts
(Sharvin contacts) and high $Z$ contacts (tunneling contacts).
\subsubsection{Low $Z$ (110) contacts}
Fig.(12) shows the conductance of a contact prepared on a (110)
face of a LSCO single crystal \cite{Dagan:2000}. The
crystal itself had $T_c$ = 33K, near optimum doping, but the local
$T_c$ at the contact was only 16K, probably due to the
manipulations used to prepare it, resulting in a local loss of
oxygen. The characteristic has the shape of an inverted V, in
agreement with the theoretical predictions for a low $Z$ (110)
surface (Fig.7). A fit to theory gives $\Delta = 5$ meV. This
result will be important when we discuss the underdoped regime
(pseudogap regime) in the next section.

\begin{figure}
\epsfxsize=3.5in \epsfbox{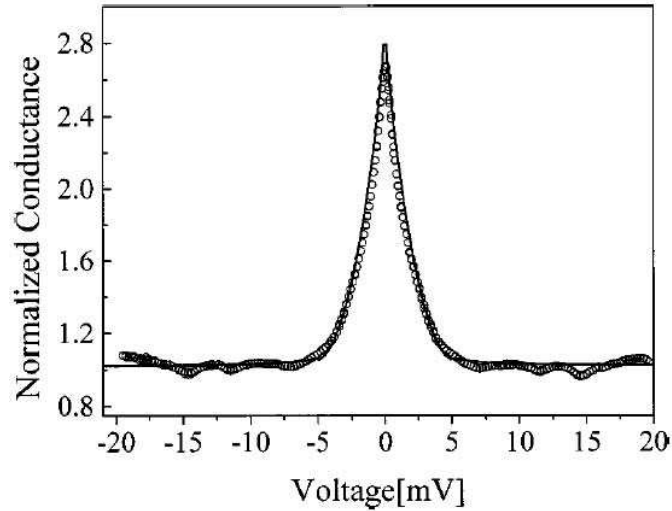}

\caption{Measured conductance characteristic of a
$Au/La_{2-x}Sr_xCuO_4$ single crystal underdoped contact. The data
was fitted for a nodal direction, giving $Z = 0.3$, and a gap of 5
meV (after \citealt{Dagan:2000}).}
\end{figure}

Low $Z$ contacts on LSCO at actual optimum doping have slightly
different characteristics. Fig.(13) shows that of a contact
obtained by electro-migration \cite{Achsaf:1996}. Upon
making contact between the tip and the sample, the resistance was
first very high, in the 100k$\Omega$ range, presumably due to an
oxygen depleted surface layer. A positive bias was then applied to
the tip, possibly attracting positively charged oxygen ions from
the bulk of the sample towards the surface. The general shape of
the characteristic is still that of an inverted V, but with a
local minimum near zero-bias. As explained in the theory section,
this local minimum is incompatible with a pure d-wave symmetry.

\begin{figure}
\epsfxsize=3.5in \epsfbox{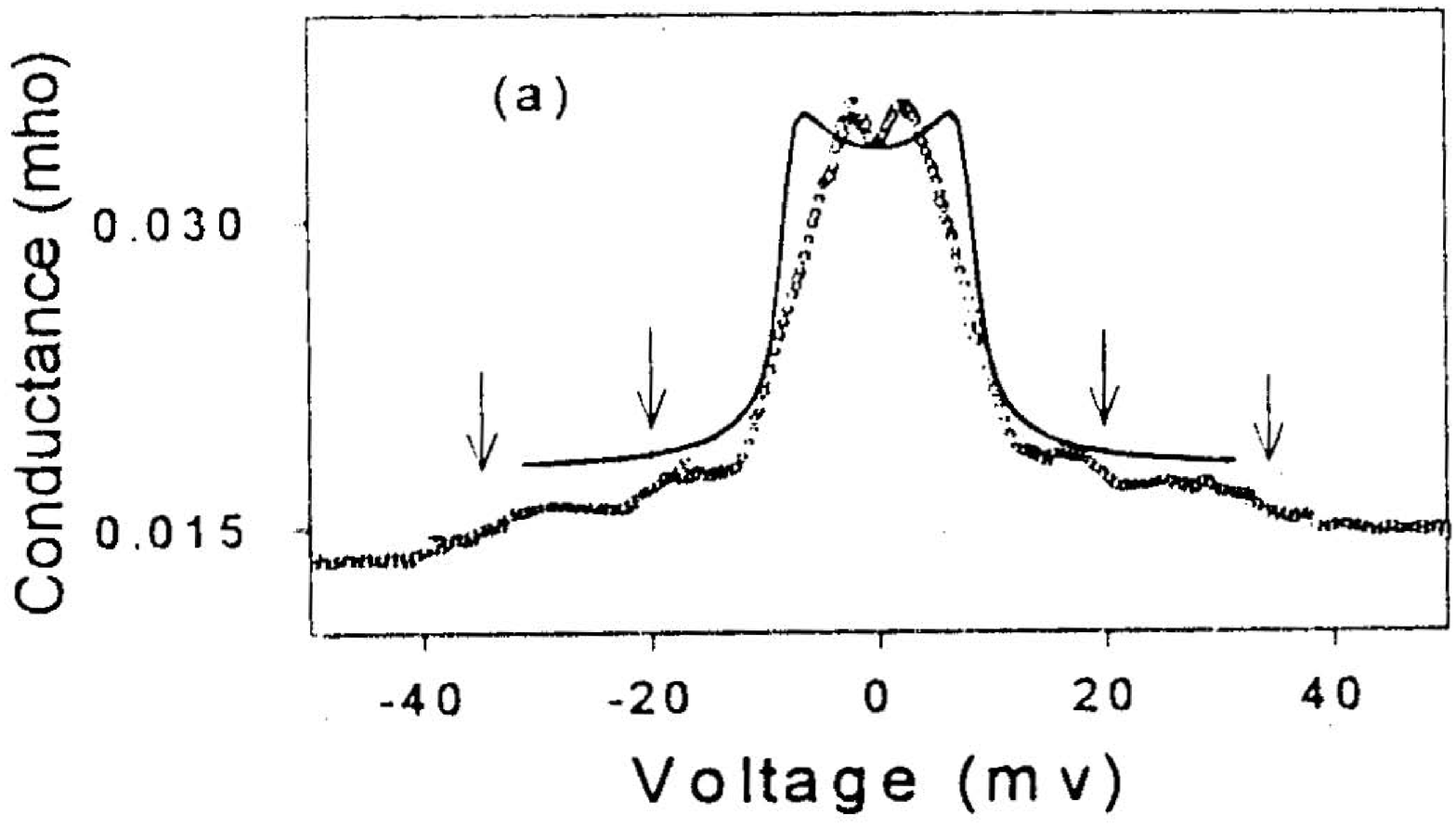}

\caption{Measured conductance characteristic of a
$Au/La_{2-x}Sr_xCuO_4$ single crystal contact near optimum doping.
Note the small split of the conductance peak at small bias. The
line is an attempt to fit the data with an s-wave gap. Arrows
indicate high bias structures possibly related to phonons. (after
\citealt{Achsaf:1996})
   }
\end{figure}

Etching the surface is another way to obtain a low $Z$ contact. On
the same crystal, this method lent characteristics having the same
general shape as that obtained on a junction prepared by
electro-migration, shown above, but with a wider separation
between the two peaks. It may be that etching exposes other
crystallographic facets besides (110), giving a characteristic
more similar to to that obtained on (100) surfaces. Similar
characteristics were obtained by Gonelli \emph{et al.} \cite{Daghero:2002} on
polycrystalline LSCO samples. They were also interpreted as
indicating a mixed symmetry.

Concerning the gap value, Achsaf \emph{et al.} \cite{Achsaf:1996} conclude
from their low $Z$ data that it is of about 9 meV in a slightly
underdoped LSCO single crystal. This is somewhat smaller than the
value of the gap obtained from tunneling contacts on the same
crystal, which is closer to 15 meV.

Relatively low $Z$ contacts ($Z$ =1) were  obtained by Wei
\emph{et al.} \cite{Wei:1998} on an optimally doped YBCO single crystal by
driving a Pt-Ir tip into the sample. The characteristics had also
the shape of an inverted V. The gap value extracted from the fit
is $27\pm4$ meV. For comparison, a regular STM measurement taken
along the c-axis on the same crystal gives a gap value of $19\pm4$
meV.
\subsubsection{High $Z$ (110) contacts}
In agreement with theory, high $Z$ contacts to (110) oriented
surfaces show a ZBCP and a weak structure at the gap edge.

Sinha and Ng \cite{Sinha:1998} have produced tunnel junctions on the edges of
BSCCO single crystals by evaporating onto them a Pb or a Ag
counter-electrode. The roughness of the edge is large (3000$\AA$),
therefore there is no well defined surface orientation. Taking
this orientation as a free fit parameter, and a smearing factor of
3.08 meV Sinha et Ng obtained a good fit of their data to KT
theory, with $Z = 2$ and $\Delta = 13$ meV. One can see from
Fig.(14) that the zero-bias conductance is larger than the normal
state conductance by a factor larger than two, and that the gap
edge has a weak signature as a dip below the normal (high bias)
conductance value followed by a progressive recovery. The $T_c$ of
the junction was 75K, presumably indicating underdoping at the
surface, either intrinsic or provoked by the contact with the
counter-electrode. The rather small gap value will be commented
upon in relation with the pseudogap issue in the next section.

\begin{figure}
\epsfxsize=3.5in \epsfbox{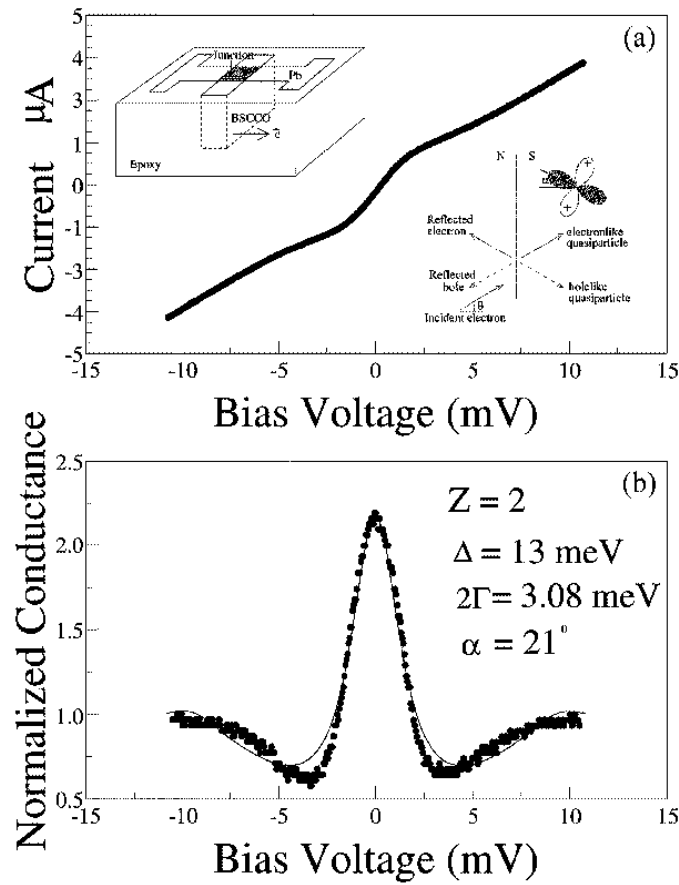} \caption{I(V) and
conductance characteristics of an in-plane contact to a
$Bi_2Sr_2CaCu_2O_8$ single crystal. Note the return to the normal
state conductance at about 10 meV (after \citealt{Sinha:1998}). }
\end{figure}

Wei \emph{et al.} \cite{Wei:1998} measured the characteristics of STM
tunnel junctions on (110) faces of an YBCO single crystal. They
show a large ZBCP, the conductance at zero-bias reaching up to 8
times the normal state value. As predicted by theory, the peak is
followed by a dip before return to the normal state (high bias)
value. The gap value obtained from the fit is $27\pm 4$ meV.

\begin{figure}
\epsfxsize=3.5in \epsfbox{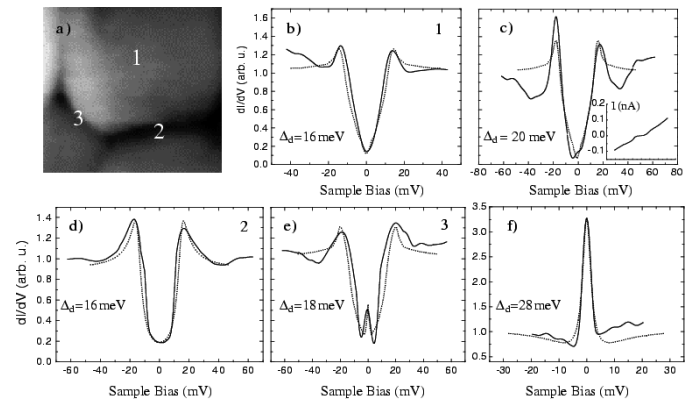} \caption{Measured
STM conductances at various positions on an YBCO grain having a
(001) oriented upper surface. The shape of the characteristics
goes from V-shape on top of the grain (b,c), to U shape near a
(100) face (d), to inverted V near a (110) face (e,f) (after
\citealt{Sharoni:2001}).}
\end{figure}

In general, the highest ZBCPs have been obtained on STM junctions.
STM reveals neatly the large anisotropy in the TDOS of the HTSC.
Sharoni \emph{et al.} \cite{Sharoni:2001} have reported observing
on the same sample, a c-axis oriented YBCO film, three types of
characteristics: respectively V-shaped on (001) areas, ZBCP
dominated on (110) edges and flat bottom on (100) edges (Fig.15)
(see also \citealt{Sharoni:2003}). These observations show that
the TDOS can vary over a length scale of the order of the
nanometer. On (110) oriented films, a ZBCP and a Gap Like Feature
(GLF) are both observed \cite{Sharoni:2002}. Macroscopic contacts
allowing in-plane tunneling into films having the (100) or (110)
orientations have been produced by a number of techniques such as:
using a Pb counter-electrode (sometimes with a thin Ag buffer
layer in order to avoid massive oxygen out-diffusion,
\citealt{Lesueur:1992}); or a copper counter electrode
\cite{Aprili:1999}. The method that we have been mostly using is
sticking a small In dot on the film's fresh surface
\cite{Krupke:1999}. These contacts are very stable and can sustain
repeated thermal cycling without damage. The exact nature of the
dielectric layer is not known. It can be the result of some loss
of oxygen at the surface, resulting in an underdoped YBCO
insulating surface; or to oxidation of the In counter-electrode by
oxygen diffusing out from the YBCO layer. The process appears to
be self-limited (this is not the case with a Pb counter-electrode,
which "pumps out" oxygen so effectively that the underlying YBCO
film can become insulating). Millimeter size contacts have typical
resistances in the convenient range of 10 to 100$\Omega$. These
junctions do not yield characteristics as ideal as those that can
be obtained by STM, apparently due to surface roughness and to the
high sensitivity of the TDOS to faceting on the nanometer scale as
demonstrated by the STM observations  of Sharoni \emph{et al.}
\cite{Sharoni:2001}. (110) facets act actually as shorts. But the
junctions have the significant advantage that they allow to take
easily measurements as a function of temperature and applied
magnetic field.

\begin{figure}
\epsfxsize=3.5in \epsfbox{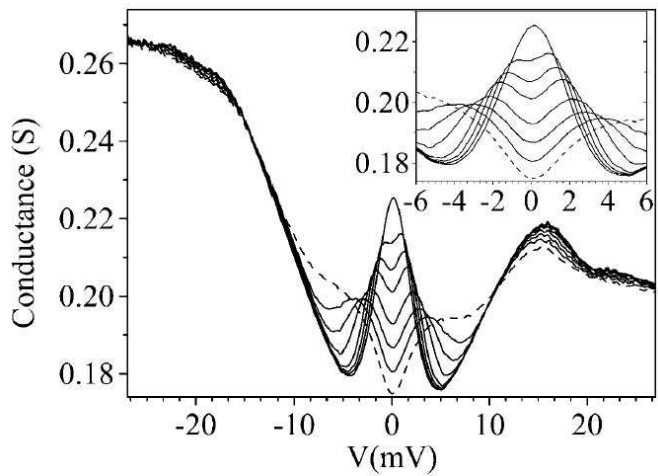} \caption{Tunneling
characteristics of an In/YBCO junction on a (110) oriented film at
increasing magnetic fields of up to 6T. Insert: measurements in
decreasing fields. (after \citealt{Dagan:2001doping}).}
\end{figure}

A typical characteristic obtained by this method on a (110)
oriented film is shown Fig.(16). Compared to STM data, the ZBCP is
considerably smeared. On the other hand, a GLF is well pronounced.
This GLF peak has been interpreted within the framework of the KT
theory as resulting from surface roughness \cite{Fogelstrom:1997},
which has about the same effect as if the surface had an effective
orientation intermediate between (100) and (110). Actually, there
is not much difference between the characteristics of macroscopic
junctions prepared on (100) and (110) oriented films. They show
similar ZBCPs and GLFs. A quantitative fit to the data taking into
account surface roughness has been presented by Fogelstrom
\emph{et al.} \cite{Fogelstrom:1997}. The GLF is well reproduced.
Its peak position is somewhat below the value of the gap.
Experimentally, its position is extremely reproducible from sample
to sample (optimally doped), and from laboratory to laboratory. It
is in fact one of the most reliable pieces of data to be found in
the HTSC literature. Its position is 17 mV for optimally doped
YBCO, with a variation of less than 1 mV between data originating
from different laboratories. From Fogelstrom \emph{et al.}
\cite{Fogelstrom:1997}, the gap value might be up to 50\% higher,
or about 25 meV, depending on the exact surface orientation
spread. Giving to the ASJ gap in optimally doped YBCO a range of
20 to 25 meV is a safe estimate.

\subsubsection {High $Z$ (110) contacts under magnetic fields}
Lesueur \emph{et al.} were the first to notice that a magnetic
field can induce a split of the ZBCP in YBCO films, and proposed
that it might be due to a Zeeman effect, under the assumption that
the ZBCP itself is due to the presence of magnetic impurities in
the vicinity of the barrier \cite{Lesueur:1992}.
Covington \emph{et al.} \cite{Covington:1997} further studied this effect, which
was given a different interpretation by Fogelstrom \emph{et
al.} \cite{Fogelstrom:1997} in terms of a Doppler shift of the energy of the ASJ
surface states. This Doppler shift is due to the superfluid
velocity corresponding to the field induced Meissner currents. The
experimental proof that the ZBCP is indeed not primarily due to
the presence of magnetic impurities, was given by Krupke and
Deutscher \cite{Krupke:1999} and by Aprili \emph{et al.} \cite{Aprili:1999}. With films
having a good in-plane orientation of the c-axis, they showed that
the ZBCP field splitting is very anisotropic, being strong when
the field is oriented perpendicular to the $CuO_2$ planes, and
undetectable when it is in the orthogonal direction (Fig.17). It
is in the first geometry that Meissner currents flow along the
CuO2 planes. If the ZBCP had been of magnetic origin, its
splitting should have been isotropic.

\begin{figure}
\epsfxsize=3.5in \epsfbox{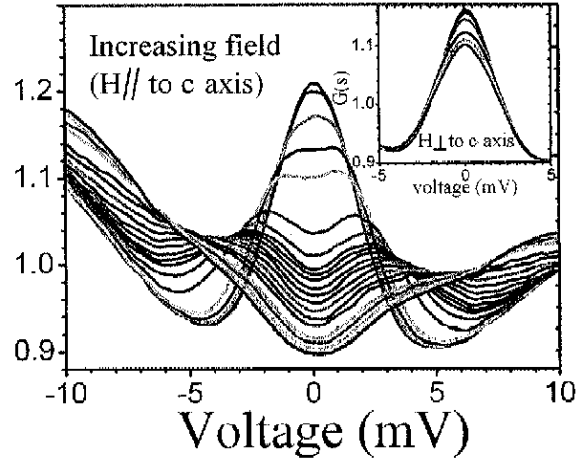} \caption{Tunneling
characteristics of an In/YBCO junction in increasing fields. The
YBCO film has the (110) orientation and the field is oriented
parallel to the surface of the film and parallel to the c-axis
(itself in-plane oriented). Note (insert) that the splitting does
not take place when the field is applied perpendicular to the
c-axis.(After R. Beck privet communication)}
\end{figure}

In the model of Fogelstrom \emph{et al.} \cite{Fogelstrom:1997},
the energy of the ASJ states is Doppler shifted by an energy equal
to $v_S\cdot p_Fcos\Theta$, where $v_S$ is the superfluid velocity
associated with the Meissner currents, $p_F$ is the Fermi momentum
and $\Theta$ is the angle that the trajectory of a tunneling
quasi-particle makes with the surface of the sample. At low
fields, theory predicts that since vS increases linearly with the
applied field, so should the ZBCP splitting, as indeed observed
experimentally (non linear effects may arise due to the presence
of a s-wave channel, \citealt{Fogelstrom:2003}). Saturation is
predicted to occur at fields of the order of the thermodynamical
critical field $H_c$, as observed by Covington \emph{et
al.}\cite{Covington:1997} (Fig.18). The strong anisotropy of the
ZBCP field splitting, and its field dependence, strongly support
the idea that d-wave symmetry is at the origin of the ZBCP. The
model of Fogelstrom \emph{et al.} \cite{Fogelstrom:1997} assumes
that vortices do not penetrate in the sample up to fields of order
$H_c$, or in other terms that there exists a strong Bean
Livingston \cite{Bean:1968} barrier. Since there is no such
barrier against vortices exit in decreasing fields
\cite{Bussieres:1976}, a strong hysteresis of the ZBCP field
splitting is expected. Again, this is in general agreement with
experiment, the splitting being larger in increasing than in
decreasing fields \cite{Krupke:1999}. In fact, according to the
Doppler shift model based on the Bean Livingston currents, there
should be no ZBCP splitting at all, or only a small one, in
decreasing fields. It is also expected that the splitting should
be strongly decreased at thickness smaller than the London
penetration depth $\lambda$, for which Meissner currents reduce as
the thickness divided by $\lambda$ . Both predictions are again in
agreement with early results obtained on films that had the (100)
orientation \cite{Krupke:1999} or the (103) orientation
\cite{Covington:1997}. For such films, the very existence of the
ZBCP is supposedly due to surface roughness. Later results,
obtained on films that did have the (110) orientation, show a more
complex behavior. In particular there is a strong splitting in
decreasing fields, and it persists even at small thickness
\cite{Dagan:2001thickness,Beck:2003}. This behavior raises
questions that will be discussed in the last section of this
review.

\begin{figure}
\epsfxsize=3.5in \epsfbox{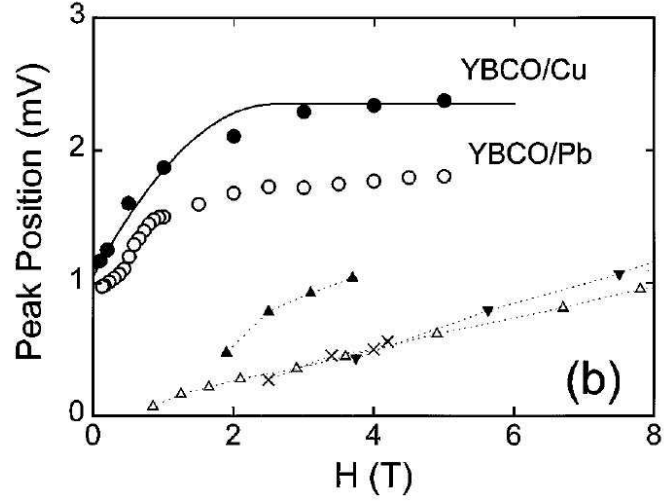} \caption{Field
dependence of the Zero Bias Conductance Peak for in-plane
junctions to YBCO films. The fit for the YBCO/Cu contact is to the
theory of Fogelstrom \emph{et al.} \cite{Fogelstrom:1997}. (after
\citealt{Covington:1997}).}
\end{figure}

\newpage
\section{ASJ SPECTROSCOPY AND THE PSEUDO-GAP ISSUE}\label{sec:V}
ASJ spectroscopy allows a good determination of the gap in the
cuprates, it gives solid evidence that they have well defined
quasi-particles and provides a good estimate of their mass
renormalization. ASJ spectroscopy is also a phase sensitive tool.
More will be said on this last topic when we discuss the
occurrence of minority components of the order parameter in the
last section of this review. It turns out that ASJ spectroscopy
also sheds light on the possible origins of the pseudogap, one of
the least understood features of the cuprates.
\subsection{Manifestations and possible origins of the pseudogap}
\label{sec:VA} There exists converging experimental evidence from
NMR spin susceptibility measurements \cite{Alloul:1989a}, heat
capacity measurements \cite{Tallon:2001}, ARPES \cite{Ding:1996},
optical measurements (see the review by Timusk and Statt,
\citealt{Timusk:1999}) and from tunneling experiments
\cite{Racah:1996,Renner:1997,Miyakawa:1997} that in underdoped
cuprates there is a loss of states at the Fermi level below a
temperature $T^*(p)$ that increases as the doping $p$ is
decreased. $T^*$ and $T_c$ have opposite variations as $p$ is
reduced below the optimum level $p_M$.  This loss of states occurs
over a certain energy range, called the pseudogap, which can be
measured by different spectroscopic methods. There are until to
day vastly conflicting views on its origin. From a
phenomenological stand point, these views belong to one of two
possible classes: either the pseudogap is a high temperature
precursor of the superconducting state, or it is strictly a normal
state property, with no direct relation to superconductivity. In
the first case, $T^*$ is the temperature below which a pairing
amplitude appears, without the phase coherence which is achieved
at the lower temperature $T_c$; it follows necessarily that $T^* >
T_c$. In the second case, $T^*$  being a normal state property is
not necessarily larger than $T_c$. In the first case, there cannot
be a crossing point between $T^*(p)$ and $T_c(p)$;  in the second
case, there may be one.

Friedel \cite{Friedel:1988,Friedel:1989} has applied the concept, developed by Mott to
describe the effects of coherent diffraction of valence electrons
from a local atomic order in liquid or amorphous metals or Hume
Rothery alloys, to the case of local 2D anti-ferromagnetic (AF)
order. The term pseudogap that he introduced describes a DOS
resembling that resulting from long range AF order, with a gap and
peaks at the gap edges, but with states within the gap and
broadened peaks.

By contrast, other authors have given to the pseudogap the meaning
of a high temperature precursor to superconductivity. Chen
\emph{et al.} \cite{Chen:2004} have investigated the regime of BCS to BE
crossover with the emphasis on finite temperature effects. They
find that in the crossover region, the order parameter is distinct
from the gap in the single particle excitation spectrum, going to
zero at $T_c$  while the gap remains finite, and goes smoothly
into the pseudogap regime above $T_c$. In the ground state, as
seen in a low temperature spectroscopic experiment, there is only
one single energy scale, the order parameter and the gap being
identical. In yet a different approach, Bernevig \emph{et al.}
\cite{Bernegiv:2003} have studied the effect of strong Coulomb effects on
superconductivity. They find that these effects generate a d-wave
gap, and at the same time reduce the superfluid density. The
stronger they are, the larger the gap value and the smaller $T_c$
will be because of Kosterlitz-Thouless effects \cite{Emery:1995}.  The transition to the insulating AF state occurs
at half filling. If Coulomb effects reduce (as may be expected in
the overdoped region), there is a smooth transition to a BCS
superconductor. In this model as well as in the BCS-BE crossover
scheme, the pseudogap is a manifestation of incipient
superconductivity.

In a different approach, Perali \emph{et al.} \cite{Perali:2000}
have studied the consequences of a strong anisotropy, both the
effective interaction and the Fermi velocity being momentum
dependent. At the anti-nodes the interaction is strong and the
Fermi velocity small, and vice versa at the nodes. One possible
mechanism of a strong momentum dependence is a charge instability
for stripe formation, occurring below a line $T^*(p)$ that starts
from a Quantum Critical Point (QCP) at $T = 0$ near optimum
doping. Other theories have emphasized the role of fluctuations
around a Quantum Critical Point (QCP), one of its manifestations
being an imaginary component of the order parameter appearing on
one side of the QCP \cite{Sachdev:2000,Ng:2004}.

Renner \emph{et al.} \cite{Renner:1997} performed STM measurements on BSCCO
single crystals at different doping levels, and observed a
pseudogap (with a substantial density of states within it) above
$T_c$, merging into the gap in the superconducting state below
$T_c$. In underdoped crystals, a pseudogap could be observed up to
room temperature (Fig.19). Even in overdoped samples, a pseudogap
was seen to persist a few 10K above $T_c$. In other terms,
$T^*(p)$ and $T_c(p)$ as measured by tunneling in BSCCO do not
cross each other. This behavior is compatible with the pseudogap
possibly being a precursor of the superconducting gap, as proposed
by the authors. An increase of the gap was also measured by break
junctions on underdoped BSCCO by Miyakawa \emph{et al.} \cite{Miyakawa:1997}.

\begin{figure}
\epsfxsize=3.5in \epsfbox{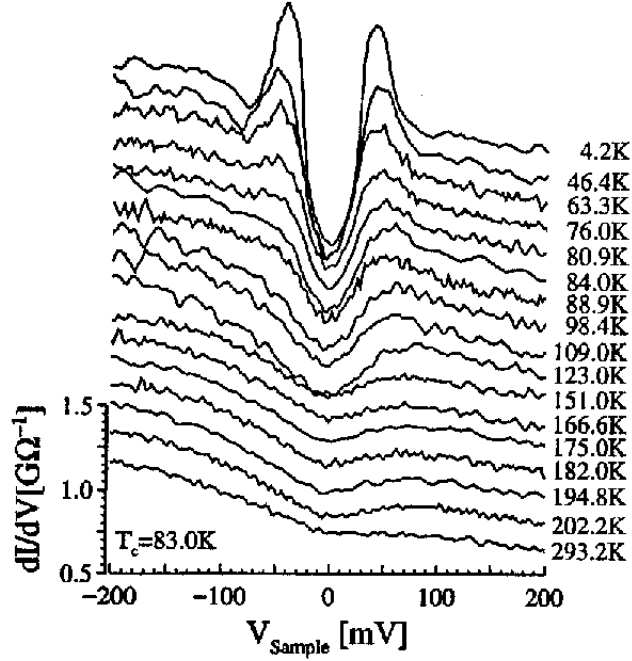} \caption{STM
characteristics measured on an underdoped $Bi_2Sr_2CaCu_2O_8$
single crystal, at temperatures ranging from 4.2K up to room
temperature. A conductance dip persists up to the highest
temperature. Note however the change in the dip amplitude when
going from 123K to 151K. (after \citealt{Renner:1997}).}
\end{figure}

On the other hand, Tallon \emph{et al.} \cite{Tallon:2001} have concluded from
an analysis of heat capacity data on a number of cuprates that
$T^*(p)$ and $T_c(p)$ do cross each other, $T^*(p)$ following a
linear behavior that extrapolates at zero temperature to a
universal "critical" concentration $p_c= 0.18$ holes/Cu. They
conclude that $T^*$ is not directly related to superconductivity.

Very recently, Alff \emph{et al.} \cite{Alff:2003} have reported that in the
electron doped compounds PrCeCuO and LaCeCuO, a pseudogap develops
only at low temperatures $T < T^* < T_c$, as can be seen by
applying magnetic fields strong enough to destroy
superconductivity, and conclude that the pseudogap cannot be a
precursor to the superconducting state. The absence of a pseudogap
opening above $T_c$  in the electron doped cuprates was also noted
by Kleefisch \emph{et al.} \cite{Kleefisch:2001}, and its presence below $T_c$
noted by Qazilbash \emph{et al.} \cite{Qazilbash:2003}.

These results (and many others that we have not quoted) do not
allow one to draw a general conclusion regarding the origin of the
pseudogap. It may be that different kinds of measurements, such as
tunneling and heat capacity, "see" different pseudogaps.
Additionally, it could also be that the pseudogap has different
origins in different cuprates. For a recent review on the
pseudogap, see Timusk and Statt \cite{Timusk:1999}.

\subsection{ASJ spectroscopy in the pseudogap regime}
\label{sec:VB}

The first important observation is that strong ASJ reflections are
observed in the pseudogap regime. This is particularly true for
ASJ bound states. ZBCPs have been observed in strongly underdoped
YBCO \cite{Dagan:2000} and BSCCO \cite{Sinha:1998}.
They persist up to $T_c$ where they vanish. The second observation
is that in the pseudogap regime gap values determined by ASJ
spectroscopy and single particle spectroscopies (tunneling or
ARPES) are different, while they roughly agree in the overdoped
regime \cite{Deutscher:1999}.

In underdoped YBCO, Yagil \emph{et al.} \cite{Yagil:1995} have reported
measurements of ASJ reflections by point contact on a-axis films
[(100) oriented surface] yielding a gap value of 13 meV. The
enhancement of the conductance below the gap is weaker than in
optimally doped YBCO. Giaever tunneling on similarly underdoped
a-axis films, with $T_c$ values in the range of 40 to 60K, gives a
gap of 40 to 50 meV \cite{Racah:1996}. (Fig.20)

\begin{figure}
\epsfxsize=3.5in \epsfbox{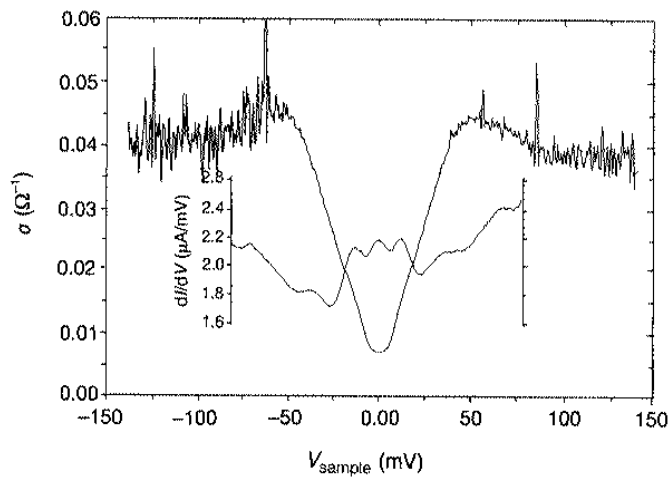} \caption{Comparison
between an ASJ (inside graph) and a Giaever characteristic
measured on similarly underdoped YBCO samples. Note the difference
in energy scales: about 15 meV for the ASJ reflection edge, and 50
meV for the Giaever gap. (after \citealt{Deutscher:1999}).}
\end{figure}

As mentioned above, in an underdoped BSCCO junction close to the
(110) orientation and having a $T_c = 70$K,  Sinha and Ng
\cite{Sinha:1998} obtained from their fit to the data an ASJ gap
of 13 meV.  This is much smaller than the 50 meV STM gap value at
similar doping \cite{Renner:1997}. Aubin \emph{et al.}
\cite{Aubin:2002} prepared tunnel junctions on optimally doped
single crystal BSCCO having a surface passivated by a thin $CaF_2$
layer. Crystals were cut and polished so as to expose (100) and
(110) oriented surfaces. On (100) surfaces, broad Giaever-like
tunneling curves were obtained with maxima at 37 meV, which is in
agreement with STM data \cite{Renner:1997}. On (110) surfaces the
conductance characteristic has the shape typical of high $Z$
junctions for that orientation, with a high ZBCP followed by a dip
before recovery to the normal state conductance. Recovery occurs
in the range of 20 to 30 meV (fig. 4a and fig.2 of Aubin \emph{et
al.}2003), which we would expect corresponds to the range of
possible gap values. A fit to this (110) data, considered as
reasonable by the authors, was nevertheless given with a gap value
of 37 meV. D'Gorno and  Kohen \cite{DGorno:1998} obtained
point-contact junctions on a nearly optimally doped (may be
slightly overdoped) BSCCO single crystal, that could be fitted
very well to a (100) orientation with a gap value of 20 meV and
$Z= 0.8$ (Fig.9). We would conclude that, for BSCCO, ASJ
spectroscopy (low $Z$ (100) contacts, moderate and high $Z$  (110)
contacts, both sets of data being dominated by ASJ reflections)
gives for BSCCO a gap of 20 to 30 meV at or near optimum doping,
and less than 15 meV in underdoped samples. By contrast, Giaever
tunneling gives gap values of 30 to 40 meV at optimum doping and
40 to 50 meV in underdoped samples. Recently, a detailed STM study
on BSCCO cleaved crystals, systematically scanning large areas of
the crystal, have shown that up to energies in the range of 20
meV, the single particle tunneling spectra are quite homogeneous
across the sample's surface even in underdoped samples
\cite{McElroy:2004}. By contrast, gap values as determined by the
bias at which the conductance is at a maximum, are quite
inhomogeneous and range from 20 meV up to 70 meV. Spectra showing
the larger gaps have weak coherence peaks. They correspond to
anti-nodal states, which are quite inhomogeneous in space. These
observations are consistent with our report of strong ASJ
reflections ranging in energies up to about 20 meV
\cite{Deutscher:1999}. This appears to be the energy scale of the
superconducting condensate. It does not increase in the underdoped
regime.

On LSCO, Gonnelli \emph{et al.} \cite{Daghero:2002} have reported measurements
of ASJ gaps that decrease in the underdoped regime, as also seen
by Dagan \emph{et al.} \cite{Dagan:2001doping}. Like in YBCO and BSCCO, Giaever
gaps do increase in underdoped samples.

The following experimental picture then emerges. Strong ASJ
reflections occur in optimally doped and overdoped samples for all
surface orientations. In underdoped samples, ASJ reflections are
weakened on (100) surfaces, but remain strong on (110) surfaces.
Gap values obtained from conductance characteristics dominated by
ASJ reflections are equal to or smaller than gap values obtained
from single particle tunneling. They decrease in the underdoped
regime, following at least qualitatively the doping dependence of
$T_c$ (Fig.21).

\begin{figure}
\epsfxsize=3.5in \epsfbox{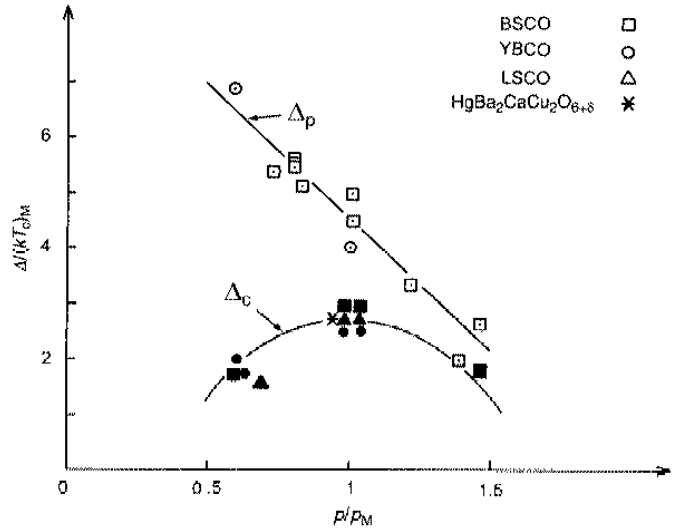} \caption{Behavior
of ASJ and Giaever (or ARPES) energy scales as a function of
doping in different cuprates. The ASJ scale ($\Delta_c$ ) follows
the same behavior as $T_c$ does, while the Giaever scale
($\Delta_p$) keeps raising as the doping is reduced (after
\citealt{Deutscher:1999}).}
\end{figure}

At low temperatures, there exists in most cuprates a tunneling
pseudogap that increases as doping is reduced. The pseudogap
appears in the normal state below a temperature $T^*$, which can
be higher or lower than $T_c$. In some cuprates, such as BSCCO,
$T^* > T_c$ in most (if not all) the doping range. In other
cuprates, the curves $T^*(p)$ and $T_c(p)$ cut each other at some
doping level $p^*$.  This is the case of YBCO and even more
clearly of the electron-doped compounds. For $p < p^*$,  the
pseudogap appears above $T_c$ and is larger than the ASJ gap. For
$p > p^*$, the ASJ gap and the Giaever gap converge. In that
regime the pseudogap does not appear above $T_c$, but in electron
doped cuprates it may be seen below $T_c$ by applying a field
strong enough to quench superconductivity.

\subsection{Compatibility of ASJ reflections and pseudogap models}
\label{VC}
\subsubsection{RVB and other models emphasizing strong correlation
effects.} A pseudogap is already implicit in the Resonating
Valence Bond (RVB) model of Anderson \cite{Anderson:1987}, which
postulates that electron correlations are the key ingredient in
the superconductivity of the cuprates. There may be however one
major difficulty with this model regarding ASJ reflections, which
concerns the nature of electronic excitations. In the RVB model,
these are not the usual quasi-particle excitations we are familiar
with in metals, but rather excitations that do not carry at the
same time charge and spin (holons and spinons). There is no
electron-hole symmetry, and it is not clear whether under these
circumstances one could at all have strong ASJ reflections, whose
very existence imply electron-hole mixing : in a Saint James
cycle, an excitation is electron-like half of the time, and
hole-like the other half.

This difficulty may be overcome in other models where strong
electron correlations play also a key role. They include
fermion-boson models with electron pockets developing near the
nodes as doping is increased (\citealt{Altman:2002}; for earlier
fermion-boson models, see
\citealt{Friedberg:1989,Ranninger:1995}). In the condensed state,
there is a bosonic field that generates a superconducting gap
\emph{a la} BCS in the electron pockets, and excitations from this
gap would have the usual electron-hole symmetry.

In another recent model originally proposed by Laughlin \cite{Laughlin:2002,Bernegiv:2003}, strong electron correlations
come on top of a conventional BCS Hamiltonian. In contrast with
the RVB model, the nature of electronic excitations is then
identical to that in ordinary metals, there are thus no problems
with ASJ reflections. Strong correlations increase continuously
the value of the BCS gap, while at the same time the superfluid
density is reduced, eventually leading to the destruction of
superconductivity as a macroscopic coherent phenomenon \cite{Emery:1995}. The (pseudo) gap appearing at high temperatures
is then a precursor of superconductivity, before it is quenched
altogether. As we shall see below, there are problems with this
part of the model.

\subsubsection{The semiconductor-superconductor and strong coupling
models} Pistolesi and Nozieres \cite{Pistolesi:2000} have calculated the
conductance of N/S contacts within two models compatible with the
existence of a gap in the DOS above $T_c$: one where it is a
normal state property, namely there is a semiconducting gap
$\Delta_o$ due to a competing order such as a Charge Density Wave
(CDW), which has no direct relation to superconductivity \cite{Nozieres:1999}; and another one, in which the pseudogap is a
manifestation of a crossover to strong coupling (Bose Einstein
limit) where the energy necessary to break a pair can be much
larger than $k_BT_c$ \cite{Leggett:1980,Nozieres:1985}.

In the normal state pseudogap scenario, ASJ reflections are
reduced by the competing order that prevents the full conversion
of incoming quasi-particles into superfluid. In the strong
coupling limit, ASJ reflections are reduced because of a mismatch
of the Fermi wave vectors when the gap approaches the Fermi level.
In both cases, there exists an energy gap $E_g$ which is a
combination of the normal state gap, or energy necessary to break
a preformed pair, and of the superconducting order parameter. In
both models, the BTK reflection coefficient $A(\varepsilon)$ is
reduced below $E_g$, this reduction becoming substantial if $E_g$
is large compared to the order parameter $\Delta$.  No strong ASJ
reflections can then occur.

In the specific example of the semiconductor-superconductor model,
the CDW order $(\Delta_0)$ induces Bragg reflections and the
superconducting order $(\Delta_m)$ ASJ  reflections. The strongest
order determines the penetration depth of the evanescent wave. If
$\Delta_0$ is larger, the specular Bragg reflection will build up
before the ASJ reflection does. The later will therefore be weak.
A similar conclusion is reached in the strong coupling limit. In
the calculated conductance characteristic, there is no structure
at the bias equal to the superconducting order parameter if
$\Delta_0 > \Delta_m$, or in the strong coupling limit.

These results are of a generic nature. They apply to different
models of the pseudogap such as the BCS to BE crossover
(\citealt{Chen:2004} or models emphasizing strong correlations
\citealt{Bernegiv:2003}, or models of competing
orders\citealt{Castellani:1997}). They are not in agreement with
the experimental findings of strong ASJ reflections, and of an ASJ
energy scale smaller than the Giaever gap, in the pseudogap
regime, \emph{if one assumes that the competing order parameter
(or the strong coupling effects) dominates all around the Fermi
surface over the superconducting order parameter.}

Note however that the calculation assumes that the normal state
gap is a full gap, namely that in the normal state there are no
states below it. This is of course an oversimplification of the
experimental situation. For PCCO, in which the normal state TDOS
has been measured at low temperatures \cite{Alff:2003},
the zero-bias conductance is still about 80\% or more of its
normal state value, so there are in fact many states below the
pseudogap. Such a high DOS in the pseudogap region is obtained in
the model of Friedel and Kohmoto \cite{Friedel:2002}. If these states are
conducting, one might expect structures in the conductance
characteristic at both biases: the superconducting order parameter
and the pseudogap scales. Two energy scales: one where the
conductance goes down, marking the ASJ gap, and one where it goes
up, marking the pseudogap, are indeed visible in the data of Yagil
\emph{et al.}\cite{Yagil:1995}. But no conductance calculations are available
for the Friedel-Kohmoto \cite{Friedel:2002} pseudogap model that we could
compare quantitatively to experiments.

An alternative approach has been tried by Pistolesi
\cite{Pistolesi:1998}, who has shown that a critical current
effect may in fact introduce in the conductance characteristic a
structure at a bias value which is typically the phase stiffness
$\Lambda$, which limits $T_c$ in the strong coupling or weak
superfluid density limit. As shown for instance by Emery and
Kivelson \cite{Emery:1995}, when the superfluid density is very
small, $T_c$ is determined by:

\begin{equation}\label{eq:e51}
kT_c=(16\pi^3)^{-1}(\Phi_0)^2(a/\lambda^2)
\end{equation}
where $\Phi_0$ is the flux quantum, $\lambda$ the in-plane London
penetration depth and $a$ a length scale equal to the coherence
length in the direction perpendicular to the CuO planes or to the
inter-plane distance, whichever is larger. Pistolesi finds that a
critical current effect will reduce the conductance at a voltage
$V_c$ given by:
\begin{equation}
\label{eq:e52} eV_c=\Lambda (k_F \xi_{phase})^{-1}
\end{equation}
where $xi_{phase}$ phase has been calculated by Pistolesi and
Strinati \cite{Pistolesi:1996} and by Marini \emph{et al.}\cite{Marini:1998}. In the vicinity
of the BCS to BE crossover, $k_Fxi_{phase}\approx 1$, and
$eVc\approx\Lambda\approx kT_c$, in agreement with experimental
results \cite{Deutscher:1999}.

\subsubsection{Two gap model}
The Rome group \cite{Perali:2000} has extended the
semiconductor-superconductor model of Pistolesi and Nozieres
\cite{Pistolesi:2000} to the case of a d-wave order parameter. In
this model, the competing order parameter dominates over the
superconducting one \emph{only in the anti-node regions}.
Specifically the origin of the pseudogap lies in the vicinity of a
Charge Density Wave (CDW)\cite{Benfatto:2000} line $T^*(p)$ , as
mentioned above, but from a phenomenological standpoint their main
results are given in terms of a gap $\Delta (\phi)$ which is
dominated by the pseudogap $\Delta_p$ near the anti-nodal points
and by the superconducting order parameter near the nodes. They
argue in favor of the weak coupling limit $\Delta < t$, where $t$
is the nearest neighbor interaction term in the tight binding
approximation, so that $T_c$ is equal within a numerical factor to
the energy scale that governs the behavior of $\Delta$ near the
nodes. In the over-doped regime, $\Delta(\phi)$ follows the d-wave
law $\Delta(\phi) = \Delta(0)cos2\phi$ over the entire angular
range. In the underdoped regime, the value of the gap at the nodal
points is uncorrelated to the (larger) characteristic energy scale
near the nodes.

This model is qualitatively in agreement with the results of ASJ
spectroscopy on (110) surfaces in the under-doped regime as
described above. The pseudogap near the anti-nodal points is
larger than the value that the superconducting order parameter
would have in the absence of this pseudogap (in the language of
Nozieres and Pistolesi, $\Delta_o > \Delta_m$ near the anti-nodal
points). Thus, these regions do not contribute to the ASJ
reflection amplitude because for the corresponding \textbf{k}
vectors normal quasiparticle reflection occurs before conversion
to the condensate takes place. Hence, the only energy scale that
will appear in ASJ spectroscopy will be that which characterizes
the angular dependence of the gap near the nodes, which is the
superconducting order parameter, itself proportional to $T_c$. The
fact that for all cuprates studied so far the ASJ spectroscopy
energy scale varies with doping as $T_c$ does, just means that
these cuprates are basically in the weak coupling limit as assumed
by the Rome group.

More generally, a momemtum dependence of the interactions leading
to the pseudogap, whatever its origin may be, seems to be a
necessity if agreement with ASJ experiments is to be achieved.
This is because these experiments tell us that there is no
pseudogap around the nodes. Theories that do not include a
momentum dependence, such as those of Chen \emph{et al.} \cite{Chen:2004},
or Bernevig \emph{et al.} \cite{Bernegiv:2003} for which the pseudogap \emph{is}
a d-wave gap that becomes the superconducting gap at low
temperatures, do not seem to be compatible with the observation of
strong ASJ reflections in the underdoped regime.

 \subsubsection{Some comments on the pseudogap}
In a previous publication \cite{Deutscher:1999} I left open the
question of the origin of the pseudogap: whether the loss of
states at the Fermi level that starts below $T^*$  is due to an
emerging pairing amplitude (a strong coupling effect), or whether
it bears no direct relation to superconductivity. Theoretical
progress in the analysis of experimental results known at this
time and new experiments reviewed in this section, and
particularly conclusions drawn from the observation of strong ASJ
reflections in the pseudogap regime, present in fact serious
difficulties for both kinds of models, at least if no momentum
dependence is included. Yet, I believe that the balance now tilts
somewhat against the preformed pairs scenario.

The formation of true bound pairs above $T_c$, in the sense of a
negative chemical potential with respect to the bottom of the
conduction band, requires a binding energy of the order of a
fraction of the band width, say of the order of the eV. Pseudo-gap
values determined by ARPES and tunneling do reach at most 10\%  of
this value. So there cannot really be bound pairs. More
specifically, for $(\Delta/EF) \approx 0.1$ as seen
experimentally, strong coupling theory \cite{Pistolesi:1996} tells
us that the two length scales $\xi_{pair}$ (the size of a pair)
and $\xi_{phase}$ (the size of a vortex core) differ only by a few
\% , and so should the corresponding energy scales. Instead, for
moderately underdoped samples ($T_c$  about half of its maximum
value), the pseudogap and the ASJ gap differ by a factor of about
4. According to this analysis, the pseudogap is not a strong
coupling effect. Additionally, there should be at low temperatures
only one energy scale in the superconducting state
\cite{Chen:2004}. In the under-doped regime, this is not the case
since the Giaever and the ASJ gaps have opposite doping
dependences (Fig.21). The recent experiments of McElroy \emph{et
al.} \cite{McElroy:2004} also establish that there is no
correlation between homogeneous, low energy excitations in the
nodal regions, and large gaps in the anti-node regions.

Ruling out the pseudogap as a \emph{homogeneous}  precursor of
superconductivity \emph{in real space and in momentum space} does
not mean that the cuprates are strictly in the BCS weak coupling
limit. In fact, they are not far from the BCS to BE crossover,
defined as shown by Pistolesi and Strinati by the condition
$(k_F\xi_{pair}) = 1$. One of the manifestations of this proximity
is the Uemura plot $T_c \propto \lambda^{-2}$ followed by all
underdoped cuprates \cite{Uemura:2002}. Another one is the
observation of strong fluctuation effects in the heat capacity
transition. These measurements allow us in fact to draw a fine
distinction between different cuprates in terms of their proximity
to the BCS-BE crossover. The heat capacity transition in YBCO can
be analyzed in terms of a mean field jump with additional
fluctuation effects \cite{Junod:1999,Marcenat:1996}. These
fluctuation effects become quite weak in overdoped samples
\cite{Junod:1999}, for which the transition becomes more and more
BCS like as doping is increased. A contrario, no mean field jump
can be identified in BSCCO, for which fluctuation effects clearly
extend several 10K above $T_c$, consistent with it being  closer
to the BCS-BE crossover than is YBCO. Enhanced fluctuations may
also reflect the more 2D nature of BSCCO. But a closer proximity
of BSCCO to the BCS-BE crossover is also consistent with
spectroscopy results. At optimum doping, the Giaever gap in BSCCO
(about 30 to 40 meV) is larger than that in YBCO (about 20 meV).
Also, the coherence length as measured by the radius of the vortex
core, is shorter in BSCCO than it is in YBCO \cite{Fischer:1998}.
These are clear indications that at comparable doping levels BSCCO
is closer than is YBCO to the condition $(k_F\xi_{pair}) = 1$.

The STM pseudogap data of Renner \emph{et al.} \cite{Renner:1997}(Fig.19) on
underdoped BSCCO reveals that there may be a difference between
characteristics measured up to 121K, and above that temperature.
The former bare indeed a strong resemblance with those measured
immediatly below $T_c$, while the later show only a weak anomaly
that remains essentially temperature independent up to room
temperature. This suggests that there might be both a
superconductivity related pseudogap between $T_c$ and 120 to 130K,
and a normal state pseudogap at higher temperatures. More data is
needed here to clarify the situation.

As for YBCO, the quasi-mean field behavior of the heat capacity at
optimum doping is fully consistent with the absence of a pseudogap
in tunneling, and with the fact that the Giaever and ASJ gaps are
identical. YBCO is in fact the only cuprate that clearly breaks
away from the Uemura plot. By overdoping it with oxygen up to
$O_7$, its superfluid density can be increased up to a factor of
two compared to its value at optimum doping, while $T_c$  remains
almost constant (it only goes down by a few degrees).
\cite{Bernhardt:1995} In the overdoped regime, YBCO presents all
the characterisitics of a strict BCS superconductor: a critical
temperature independent from the superfluid density, a sharp heat
capacity transition, identity between the Giaever and ASJ gaps. It
would be really surprising if nearly optimum doped BSCCO, not so
different after all, would show precursor effects of
superconductivity up to room temperature. May be such strong
precursors effects could be found in more strongly underdoped
cuprates, if the condition ($k_F\xi_{pair}$) could be reached.
Research is still going on in this area.
\newpage
\section{SYMMETRY STUDIES AND SPIN EFFECTS}\label{sec:VI}
In this last section, I would like to mention two topics of
current interest in ASJ spectroscopy: effects on the symmetry of
the order parameter in the cuprates of different perturbations
such as non-optimum doping, applied magnetic fields and proximity
with a normal metal. A few words on spin effects are added at the
end of this section.

Phase sensitive experiments are the only ones that can lead to
definite conclusions regarding the symmetry of the order
parameter. Such were the corner SQUID experiments of Wollman
\emph{et al.},\cite{Wollman:1993} and those of Tsuei and Kirtley
\cite{Tsuei:2000pairing,Tsuei:2000phase} that have established
that the order parameter in the cuprates has a dominant d-wave
symmetry. These experiments have however left two interesting
questions unanswered. First, because they measure phase
differences of the order parameter at surfaces or interfaces
(grain boundaries), they are not sensitive to possible changes of
the symmetry between the surface and the bulk. Muller has recently
raised this issue, and has argued that a s-wave channel may in
fact dominate in the bulk \cite{Muller:2004}. Second, they are not
well suited to detect the existence of a small imaginary minority
component. In the experiments of Tsuei \emph{et al.} for instance,
the signature of pure d-wave symmetry is a spontaneous half flux
quantum at a tri-crystal junction. A small $id_{xy}$ component
would only slightly modify this flux value, and will be undetected
if this change is smaller than the margin of error in the measured
value. Additionally, these experiments were evidently not designed
to study possible effects of strong applied fields on the symmetry
of the order parameter. ASJ spectroscopy is ideally suited to
answer such questions. ASJ bound states are very sensitive to the
existence of a small imaginary component of the order parameter,
which has the immediate effect of removing the nodes. And ASJ
spectroscopy can be easily performed under applied fields.

The formalism of Kashiwaya and Tanaka \cite{Kashiwaya:1995} can be used to
calculate I(V) characteristics for any order parameter
$\Delta(\phi)$. For instance for:
\begin{equation}\label{eq:e61}
\Delta(\phi)=\Delta_0cos2\phi+i\Delta_1sin2\phi
\end{equation}
(the $d+id$ symmetry), the (110) ZBCP is split (peak to peak) by
$\Delta_1$. The same holds for a ($d + is$) symmetry. Such
symmetries have been discussed by Yang and Hu \cite{Hu:1994}. These order
parameters break time reversal symmetry, implying the flow of
boundary currents \cite{Laughlin:1998}.
\subsection{ASJ bound states under applied fields}
\label{sec:VIA}
 The effect of Meissner screening currents also
leads to a split of the ZBCP as discussed in section IV. When a
ZBCP splits under an applied field, how can we know whether it
does so because of field induced Meissner currents, or because it
is the field itself (not the currents) that has induced a change
in the symmetry of the order parameter, as proposed for instance
by Laughlin \cite{Laughlin:1998}?

Beck \emph{et al.} \cite{Beck:2004} have remarked that the current effect
should cancel out in \emph{decreasing fields}, because it is known
that there is no Bean-Livingston surface barrier against flux
\emph{exit}. A ZBCP split in decreasing fields, if observed,
should therefore be primarily a field effect, and not a current
effect. They have measured such a split in the characteristics of
junctions prepared on (110) oriented YBCO films, and have shown
that it follows the law:
\begin{equation}\label{eq:e62}
\delta(B)=AB^{1/2}
\end{equation}
where $2\delta$ is the ZBCP split expressed in meV, $A = 1.1$
meV/$T^{1/2}$ and $B$ is the field in Tesla. This law was seen to
hold up to a field of 16T (Fig.22).It is in agreement with a
prediction by Laughlin \cite{Laughlin:1998} regarding the
amplitude $\Delta_1$ of a field induced $idxy$ component. Node
removal costs an energy proportional to $|\Delta_1|^3$, but an
energy proportional to ($B\cdot\Delta_1$) is gained because of the
magnetic interaction with the field of the moment produced by the
\emph{circulating} currents proportional to the amplitude of the
$id_{xy}$ component. Minimization of the sum of the two terms
leads to a law of the form \ref{eq:e62}. The value of the
coefficient $A$ found experimentally is in quantitative agreement
with theory.

\begin{figure}
\epsfxsize=3.5in \epsfbox{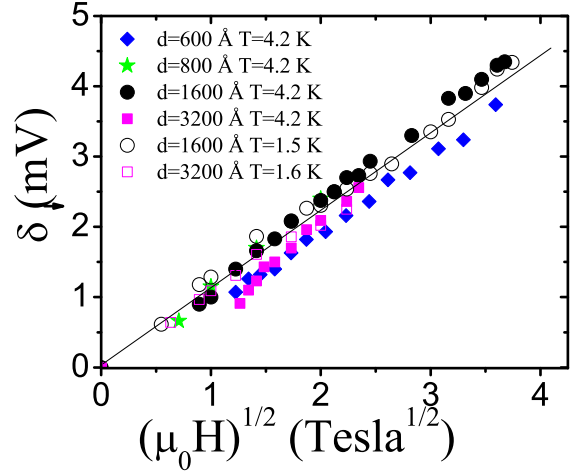} \caption{Split of
the ZBCP in a number of In/YBCO junctions on (110) oriented films,
of different thickness, measured in decreasing fields. The split
follows a square root law, with a coefficient of the order of 1
mV/$T^{1/2}$. (after \citealt{Beck:2004}) }
\end{figure}

\subsection{Doping effect on the symmetry}\label{sec:VIB}
Covington \emph{et al.} \cite{Covington:1997} have reported a
spontaneous split of the ZBCP in YBCO in-plane tunneling, and have
interpreted it as an effect of spontaneous time reversal symmetry
breaking. Fogelstrom \emph{et al.} \cite{Fogelstrom:1997}have
proposed that this spontaneous split results from the emergence at
the surface of an \emph{is} component of the order parameter, an
emergence made possible by the local depression of the main
\emph{d} component.

Dagan \emph{et al.} \cite{Dagan:2001doping} have reported that in YBCO the
spontaneous ZBCP split occurs only in overdoped samples, where it
follows the law:
\begin{equation}\label{eq:e63}
\delta = C (p - p_M)
\end{equation}
where $p$ is the doping level and $p_M$ its optimum value (at
maximum $T_c$) (Fig.23). These results were obtained on oxygen
overdoped films. A similar law was reported by Sharoni \emph{et
al.} \cite{Sharoni:2002} on Ca overdoped YBCO films.

\begin{figure}
\epsfxsize=3.5in \epsfbox{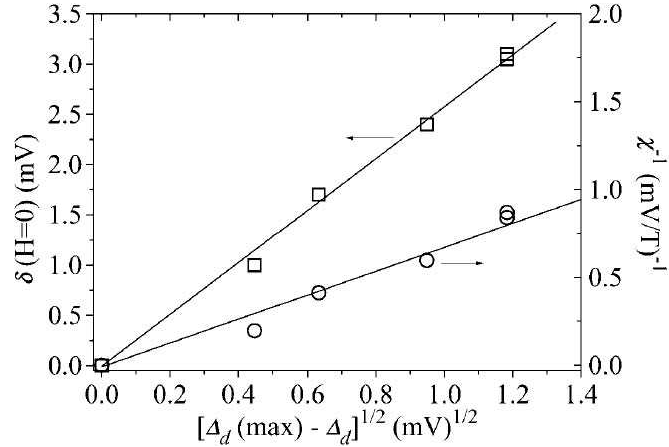} \caption{Dependence
of the spontaneous ZBCP split in In/YBCO junctions on (110)
oriented films, as a function of a parameter proportional to the
doping level. Overdoping was achieved by increasing the oxygen
content. A spontaneous split is only found in overdoped samples.
The inverse of the initial slope of the ZBCP field splitting, or
susceptibility, is also shown. The susceptibility diverges near
optimum doping. The behavior of the spontaneous splitting and of
the susceptibility are indicative of the presence of a quantum
critical point near optimum doping, beyond which the order
parameter develops a small imaginary component (after
\citealt{Dagan:2001doping}).}
\end{figure}

There is no direct way to know from tunneling measurements whether
the spontaneous imaginary component supposedly responsible for the
ZBCP split has the $s$ or the $d_{xy}$ symmetry. An $s$ symmetry
would mean that the strength of the sub-dominant s-channel,
emerging at the surface as postulated by Fogelstrom et al, \cite{Fogelstrom:2003},
has a strong doping dependence. Alternatively, the imaginary
component might be a bulk property. Friedel and Kohmoto \cite{Friedel:2002}
have predicted that an idxy component should appear in the
ovedoped regime, while the symmetry is pure d-wave in the
underdoped one, as reported by Dagan \emph{et al.} \cite{Dagan:2001doping}. In
their theory, the d-wave symmetry does not come about because of
the interaction responsible for pairing , but rather due to the
symmetry of the carriers wave function. Yet another possibility is
that the change of symmetry at optimum doping reflects the
existence of a quantum critical point \cite{Dagan:2001doping,Sachdev:2000}.

Some ASJ data indicating a possible change of symmetry near
optimum doping are also available on LSCO
\cite{Dagan:2000,Achsaf:1996} and on the electron doped PCCO
\cite{Qazilbash:2003}. The electron doped cuprates had long been
considered an exception to the d-wave symmetry
\cite{Fournier:1998}, inter alia because of the absence of a ZBCP
\cite{Alff:1997}. However, more recent tri-crystal experiments
,\cite{Tsuei:2000pairing,Tsuei:2000phase}, indicated a d-wave
symmetry. But very recently, a change of behavior of PCCO has been
reported as a function of doping, including the ASJ data of
Qazilbash \cite{Qazilbash:2003} already mentioned and penetration
depth data in overdoped PCCO is better fitted by a nodeless order
parameter \cite{Skinta:2002}. It could well be that the long
standing controversy on the symmetry of the order parameter in the
electron doped cuprates is on its way towards a resolution in
terms of a doping dependence. Electron doped cuprates are
naturally overdoped, which may explain the early results pointing
to a nodeless behavior.

A doping dependent symmetry, if confirmed as a bulk property,
would be of some consequence for our understanding of the HTS
mechanism. Many of the proposed theoretical models, such as
($t.J$) models, give a prominent role to the anti-ferromagnetic
coupling parameter $J$, with this interaction being the primary
coupling channel.  A pure d-wave symmetry follows necessarily in
such models, at any doping level.  On the contrary, in the model
of Friedel and Kohmoto, the d-wave symmetry does not follow from
the pairing interaction itself, but rather from the symmetry of
the electronic wave functions, as they are affected by the
proximity of the AF state. In that case, Friedel and Kohmoto show
that the order parameter symmetry changes with doping, and in a
manner that fits the experimental observations of Dagan \emph{et
al.} \cite{Dagan:2001doping}, an imaginary component appearing
beyond optimum doping. In fact, it has been claimed very recently
on the basis of high temperature expansions in the thermodynamical
limit that the one band ($t,J$) model does not lead to a
superconducting state \cite{Pryadko:2004}, contrary to what had
been proposed earlier from numerical work on finite size systems.
So it may be that more conventional interactions such as the
electron-phonon interaction will now receive renewed attention.

\subsection{Proximity effect on the symmetry}\label{sec:VIC}
Wei \emph{et al.} \cite{Wei:1998} noted the possibility of a
proximity effect between a normal tip and YBCO resulting in a
partial s-wave character of the order parameter in the later.
Gonnelli \emph{et al.}\cite{Daghero:2002} studied low $Z$ contacts
($Z<0.5$) on LSCO samples as a function of Sr doping. They
analyzed the conductance characteristics in terms of a complex
order parameter and determined the intensity of each component as
a function of doping. They found them to be of the same order.
This is in sharp contrast with results obtained on high $Z$
contacts with YBCO, described in the above sub-section. There, the
$is$ (or $id$) component is never more than a fraction of the
dominant $d$ component. Kohen \emph{et al.} \cite{Kohen:2003} have
recently reported a systematic study of the intensity of the
minority component in YBCO as a function of the barrier
transparency, for contacts point contact having all $Z < 1$
(fig.24). They conclude that the value of $Z$ has a strong
influence on the intensity of this component, which according to
their analysis has the $is$ symmetry. At low $Z$ values, both
components have similar values. The is component diminishes
quickly for $Z > 0.5$. The authors interpret their results in
terms of a proximity effect. They argue that a good contact with
the normal metal depresses the d-wave order parameter near the
interface, because this symmetry channel is very unfavorable for a
proximity induced order parameter in N. In the presence of a
sub-dominant s-symmetry channel in S, that symmetry can then
manifest itself near the interface. An interesting observation by
Kohen \emph{et al.} \cite{Kohen:2003} is that the excitation gap
for the case of a ($d + is$) order parameter, $\Delta_g =
(\Delta_d^2 + \Delta_s^2)^{1/2}$, does scale with $T_c$, meaning
that it retains at the surface its bulk value, although the
respective weights of the two components vary with $Z$ from
junction to junction.

\begin{figure}
\epsfxsize=3.5in \epsfbox{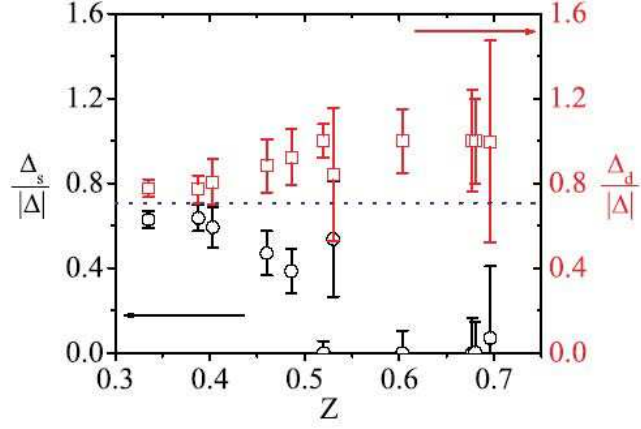} \caption{Variation
with the barrier parameter $Z$ of the d-wave component and of the
is-wave component at $Au/YBCO$ contacts fitted to the (100)
orientation. The \emph{is} component becomes of the order of the
d-wave one for high transparency barriers, suggesting that it is
due to a proximity effect. (after \citealt{Kohen:2003}).}
\end{figure}

\subsection {Spin effects}\label{sec:VID}
ASJ reflections are profoundly modified at the interface with a
ferromagnetic metal. This topic has been recently reviewed by
Zutic \cite{Zutic:2004}. As the spin polarization increases, the
conductance of a Sharvin contact at subgap voltages decreases:
spin conservation requires that the ASJ reflected hole must have a
spin opposite to that of the incoming electron, a process
incompatible with full spin polarization in the ferromagnet
\cite{deJong:1995}. This has been verified experimentally
\cite{Upadhyay:1998,Soulen:1998}. Zutic and Valls
\cite{Zutic:1999} and Zutic and Valls \cite{Zutic:2000} have
however pointed out that this low bias conductance decrease as the
polarization is increased is only a general property for contacts
where the Fermi velocities of both sides are nearly matched. When
they are not, the zero-bias conductance may in fact initially rise
with the polarization. Care should therefore be exercised when one
attempts to extract the value of the polarization from conductance
curves. Using for the fit the BTK parameter $Z$, that does not
distinguish between the effect of a dielectric barrier and that of
a Fermi velocity mismatch, may not be justified. Chen \emph{et
al.} \cite{Chen:2001} have studied transport across the interface
between an YBCO layer and a high spin polarized oxide. They
conclude that spin polarization tends to diminish the ZBCP
feature. There has been recently experimental interest in spin
injection from ferromagnets into High-$T_c$ cuprates
\cite{Dong:1997,Vasko:1997,Fu:2002}. Ngai \emph{et al.}
\cite{Ngai:2004} have developed an injection scheme allowing a
simultaneous STM measurement, and have shown that spin injection
reduces and broadens the ZBCP feature. Theoretical treatment has
taken into account the influence of ASJ reflections at the
interface, including the effect of d-wave symmetry and
particularly that of surface bound states
\cite{Zhu:1999,Kashiwaya:1999,Merril:1999,Zutic:1999}.

Mesoscopic studies of ASJ reflections cover a large field that
could be the subject of a review all by itself. We limit ourselves
here to a particular situation that is drawing increasing
attention, that where two ferromagnetic tips in close proximity
are in contact with a superconductor, the distance between them
being shorter than the coherence length. Assume that the tips are
fully polarized, that they are connected to a busbar and that a
difference of potential is applied between this busbar and the
superconductor. Then an electron coming from one the ferromagnetic
legs cannot be ASJ reflected as a hole in that same leg. It can
however be reflected in the other leg, provided the polarizations
in the two legs are antiparallel \cite{Deutscher_Feinberg:2000}.
As a result, the resistance of the device will depend on the
relative magnetic polarizations of the two legs: it will be high
if they are parallel, low if they are antiparallel. One can
consider the device as some kind of transistor: the resistance can
be modified by applying a local magnetic field that can reverse
the polarization of one of the two legs. More fundamentally, in
the case of antiparallel polarizations, the incoming electron in
one leg and the reflected hole in the other one can be considered
as two electrons of the same Cooper pair separated in space, a
situation that can have interesting implications
\cite{Recher:2003}. The basic prediction of a resistance sensitive
to the relative polarizations has just been verified
experimentally, as well as the exponential decay of the effect on
the scale of the coherence length \cite{Beckmann:2004}. The
amplitude of the effect is sensitive to the exact geometry and to
scattering inside the superconductor \cite{Melin:2002}.
\newpage
\section{CONCLUSIONS}
ASJ reflections are a powerful tool for the study of the nature of
electronic excitations in superconductors, determination of energy
gap values and studies of the symmetry of the order parameter.
Strong ASJ reflections have been observed in all cuprates tested
so far, including in the underdoped- pseudogap regime. Because
their occurrence implies electron-hole mixing, it follows that the
nature of electronic excitations in the cuprates is similar to
that in ordinary metals. This is an important result, which can
help to discriminate between the predictions of different
theoretical HTSC models. From the energy dependence of ASJ
reflections, one can infer that the scale of coherent, homogeneous
superconductivity is on the order of 20 meV. Pseudogap values
substantially exceeding this value may not be related directly to
superconductivity, and are apparently characteristic of the
anti-node directions. If the pseudogap is a high temperature
precursor of superconductivity, it must be strongly momentum
dependent. The same holds if it is the manifestation of a
competing order. ASJ doping dependence gives indications for the
existence of an additional interaction channel, besides the one
giving rise to d-wave symmetry. This additional channel becomes
stronger as doping is increased beyond optimum doping. ASJ
spectroscopy under applied magnetic fields provides a tool for the
study of surface currents, including Meissner currents due to an
effective Bean-Livingston barrier against vortex penetration, and
currents possibly linked to an imaginary component of the order
parameter induced by the magnetic field.

\section*{Acknowledgments}

I am greatly indebted to Philippe Nozieres and to Fabio Pistolesi
for numerous illuminating discussions and for giving me access to
their unpublished results on Andreev - Saint James reflections in
the strong coupling regime. Many thanks are due to Jacques Friedel
for his patient explanations on the origin of the pseudogap
concept, and to Roger Maynard for many fruitful discussions on ASJ
reflections and related topics. Much help from and discussions
with Meir Weger, Bernard Raveau and Alex Revcolevschi are
gratefully acknowledged. A special thank is due to Amir Kohen and
to Roy Beck for providing a number of useful numerical
simulations. The major part of this review was prepared while I
was a guest at the Institut Laue - Langevin, which I would like to
thank for its support. I would also like to thank warmly Efim Kats
for his hospitality during this stay, and for many interesting
discussions. Finally, I would like to thank Yoram Dagan, Amir
Kohen, Roy Beck and Guy Leibovitch of the Tel Aviv team, and Oded
Millo and Gad Koren for their many contributions to this work.
Support from the Israel Science Foundation, from the Heinrich
Hertz - Minerva Center for High Temperature Superconductivity and
from the Oren Family Chair for Experimental Solid State Physics is
gratefully acknowledged.
\newpage

\bibliographystyle{apsrmp}

%\bibliography{nc,cds,sw,connes}

\bibliography{ASJDeutscher}

%\centerline{FIGURE CAPTIONS}

\end{document}